\documentclass[prd,12 pt,tightenlines, nofootinbib, superscriptaddress]{revtex4}
\usepackage{amsfonts,amssymb,amsthm,bbm,amsmath,eufrak}
\usepackage{verbatim}
\usepackage{graphicx}
\usepackage{subfig}
\usepackage{array}
\usepackage{tikz}
\usepackage{hyperref}
\newcommand\numberthis{\addtocounter{equation}{1}\tag{\theequation}}
\newcommand{\C}{{\mathbb C}}
\newcommand{\N}{{\mathbb N}}

\newcommand{\Z}{{\mathbb Z}}

\newcommand{\one}{\mathbbm{1}}

\newcommand{\cH}{{\mathcal H}}

\newcommand{\id}{\mathbbm{1}}

\newcommand{\be}{\begin{equation}}
\newcommand{\ee}{\end{equation}}
\newcommand{\beq}{\begin{eqnarray}}
\newcommand{\eeq}{\end{eqnarray}}
\newcommand{\bea}{\begin{eqnarray}}
\newcommand{\eea}{\end{eqnarray}}

\newcommand{\bs}{\begin{split}}
\newcommand{\es}{\end{split}}

\newcommand{\bra}{\langle}
\newcommand{\ket}{\rangle}

\newcommand{\rd}{\mathrm{d}}

\newcommand{\bpm}{\begin{pmatrix}}
\newcommand{\epm}{\end{pmatrix}}
\newcommand{\bvm}{\begin{vmatrix}}
\newcommand{\evm}{\end{vmatrix}}

\begin{document}
\title{Bulk amplitude and degree of divergence in 4d spin foams}
\author{Lin-Qing Chen}\email{lchen@pitp.ca}
\affiliation{Perimeter Institute for Theoretical Physics, Waterloo, Ontario, Canada}
\affiliation{Department of Physics, University of Waterloo, Waterloo, Ontario, Canada}

\begin{abstract}
We study the 4-d holomorphic Spin Foam amplitude on arbitrary connected 2-complexes and degrees of divergence. With recently developed tools and truncation scheme, we derive a formula for a certain class of graphs, which allows us to write down the value of bulk amplitudes simply based on graph properties.  We then generalize the result to arbitrary connected 2-complexes and extract a simple expression for the degree of divergence only in terms of combinatorial properties and topological invariants. The distinct behaviors of the model in different regions of parameter space signal phase transitions. In the regime which is of physical interest for recovering diffeomorphsim symmetry in the continuum limit, the most divergent configurations are melonic graphs.  We end with a discussion of physical implications.
\end{abstract}

\maketitle

\tableofcontents

\section{Introduction}

%
%
%
%
%
%
%

Understanding the behavior of divergence is a crucial aspect of studying quantum field theory. Ultraviolet divergence comes from integrating degrees of freedom to arbitrarily high energy scales and should be removed through renormalization, while infrared divergence is due to an infinite number of soft massless particles and does not influence measurable quantities. 
Spin foam models attempt to define a path integral formalism of Quantum Gravity. The generic divergences in them are very different to those in quantum field theory.  In spin foams, the existence of minimum length scale removes the ultraviolet divergence \cite{Rovelli:2011eq, Perez:2012wv, Rovelli:2014ssa}.  However, the arbitrarily large length scale degrees of freedom in the path integral can in principle lead to infrared divergences.  The divergences from self-energy and radiative corrections in 4-d models have been studied in \cite{Perini:2008pd, Krajewski:2010yq, Riello:2013bzw}. The general structure has been studied in 3-d \cite{Matteo1, Matteo2, Freidel:2009hd, Carrozza:2011jn, Baratin:2014yra} and also in 4-d topological group field theory \cite{Carrozza:2012kt, Baratin:2013rja}.  Understanding the general behavior of divergences is a basis for studying renormalization in spin foams.

The divergences in spin foam models (with vanishing cosmological constant) are expected to encode information about gauge symmetry \cite{Freidel:2002dw, Baratin:2006gy, Bonzom:2013ofa, Dittrich:2011vz, Dittrich:2014rha}.   Diffeomorphism symmetry leads to non-compact gauge orbits, thus a path integral over such orbits leads to divergence.  This is known exactly in the 3-d Ponzano-Regge Model.  Residual action of the diffeomorphism group acts at the vertices of triangulation of a 3-d manifold as vertex translation symmetry.  In the Ponzano-Regge Model, the divergences which are due to this translation symmetry can be removed by using the Faddeev-Poppov procedure \cite{Freidel:2002dw}.  3-d gravity is topological and has no local degrees of freedom, hence its continuum limit is fully described by the discrete model, which is not the case for 4-d gravity.  In the case of 4-d models, the situation is non-trivial, because the diffeomorphism symmetry in discrete models is broken\cite{Dittrich:2008pw, Bahr:2009ku} and is only expected to be recovered in the continuum limit through renormalization \cite{Oeckl:2004yf, Oeckl:2002ia, Bahr:2011uj, Dittrich:2012jq, Bahr:2014qza, Dittrich:2014ala}.  Nevertheless, the discretized 4-d Regge action has the vertex translation symmetry under the 1--5 refining Pachner Move \cite{Bahr:2009ku}.  Therefore, understanding the behavior of divergence in 4-d spin foam models can give us hints about the residue of diffeomorphisms present in the model.

As a spin foam model is defined by a path integral over all the geometrical degrees of freedom with a given boundary, it is crucial to understand what type of geometry has the dominant contribution to the partition function. It will give us hints as to whether the theory will lead to smooth 4-d geometry in its continuum limit.  In the colored tensor models \cite{Gurau:2010ba, Gurau:2011xq, Gurau:2011aq,  Bonzom:2011zz, Gurau:2013cbh}, as it has been shown in the 1/N expansion, the dominant contribution to the partition function comes from melonic graphs, which leads to branched polymers phase \cite{Gurau:2013cbh, Gabrielli:1997zy}. Such a complete analysis of dominant contributions has always been missing in non-topological 4-d spin foam models.   


This paper aims at providing the answer to the above questions.  The quantity of interest is the \emph{bulk amplitude}, which is the evaluation of partition function on a fully contracted 2-complex,  or the evaluation on a connected 2-complex with zero boundary spins.  We obtain a very good approximation to a 4-d spin foam bulk amplitude on an arbitrary 2-complex. This allows us to derive the exact degree of divergence for any amplitude for the first time.

The model we use to obtain these results is a 4-d Riemmanian spin foam model with vanishing cosmological constant \cite{Banburski:2014cwa, Banburski:2015kmc}. It arose from writing spin foams in terms of coherent states using the SU(2) holomorphic representation beginning with the work in \cite{coh1, Livine:2007ya} and a lot of recent development in \cite{Freidel:2009nu, Freidel:2009ck,  Freidel:2010tt,  Dupuis:2010iq, Borja:2010rc, Dupuis:2011fz, Livine:2011gp, Dupuis:2011wy,  Dupuis:2012vp, Freidel:2012ji, Freidel:2013fia, Hnybida:2014mwa}.  In the holomorphic representation complicated integrals over SU(2) group elements can be rewritten as spinor integrals over the complex plane. This allows for exact evaluations of complicated spin network functions \cite{Freidel:2012ji, Hnybida:2014mwa}.  With an alternative way of imposing the simplicity constraints, the model still has the same semiclassical limit as the EPRL/FK model \cite{ Conrady:2008mk, Barrett:2009gg, Han:2011rf} at the leading order \cite{Banburski:2015kmc}, but the computability has been tremendously simplified. Using this model, we were able to evaluate 4-d Pachner moves \cite{Banburski:2014cwa}, and in this work we study the dynamics for arbitrary many simplices.  Even though the model we are studying has Riemmanian signature, its advantage of computability allows us to extract some information about its dominiant degrees of freedom, symmetry and hints of continuum limit, which will be beneficial for the Lorentzian case.

The structure of this paper is as follows:  In the section \ref{Brief Review of Holomorphic Spin Foam Model}, first we briefly review the holomorphic representation and the spin foam model.  We then go on to analyze the structure of generic 2-complexes and introduce the concept of an optimal spanning tree in \ref{Analyzing the graph structure}.  Sections \ref{pgf}, \ref{homogeneity map},\ref{loop identity and the truncation} serve as a review of the techniques and methods of evaluating partition functions which we have developed in the previous work \cite{Banburski:2014cwa}, and which we generalize to arbitrary 2-complexes. In the next two sections \ref{Truncated bulk amplitude} and \ref{Degree of divergence} we obtain the main results of the paper.  For a certain class of graphs which contain optimal spanning trees we derive a formula which allows us to write down a good approximation to the bulk amplitude simply based on the combinatorial properties of the 2-complex. We also introduce \emph{simplifed diagrams} to make the structure of the result more transparent.  We then generalize the result to arbitrary connected 2-complexes and extract a simple expression for the exact degree of divergence of the bulk amplitude only in terms of the number of vertices, faces and edges:
\be
D(\mathcal{G}) = \Lambda^{(\eta +2) |F| - 6|E| +3|V| - 3},
\ee
where $\Lambda$ is a large spin cut-off and $\eta$ is the power of face weight $(2j+1)^\eta$.   When $\eta =3$, the 5--1 Pachner move has $\Lambda^2$ divergence, while 4--2 Pachner move is finite. This is the desired degree of divergence one would expect for recovery of diffeomorphism symmetry in the model \cite{Baratin:2006gy, Dittrich:2011vz}.  The above expression can also be written in terms of topological invariants. In terms of a topological quantity called the \emph{degree of the graph} $\omega_{4d}(\mathcal{G})$, we find 
\be
D(\mathcal{G}) = \Lambda ^{(\eta-2)  |F| - 4 \omega_{4d}(\mathcal{G}) /3+13}.
\ee
Hence there are regimes of the face weight $\eta$ with distinct behaviors and we analyze them in section \ref{degree}. We find that when $\eta =2$, the divergence solely depends on the degree of the graph as in the colored tensor models \cite{Gurau:2011xp}. In this case the continuum limit defined by summation of all graphs is described by the branched polymers phase\cite{Gurau:2013cbh}.  In the region when 5--1 move is divergent while 4--2 is not, the dominant contribution comes from melonic graphs while maximizing the number of vertices. We finish with discussing the physical implications for different scenarios of defining the continuum limit, and point towards a few possible resolutions such that the model is not peaked on the non-geometrical configurations.


\section{Preliminaries}
\label{Preliminaries}
In this section we briefly review the holomorphic spin foam model and collect results that will be necessary for deriving the expression for the truncated bulk amplitude on arbitrary 2-complexes.
\subsection{Review of the holomorphic spin foam model}
\label{Brief Review of Holomorphic Spin Foam Model}
In this paper, we use a bra-ket notation for a spinor and its conjugate, due to the convenience of expressing contractions of spinors:
\be
  |z\ket = \bpm z_0 \\ z_1 \epm, \qquad  |\check{z}\ket \equiv  |z] = \bpm -\overline{z}_1 \\ \overline{z}_0 \epm, \qquad    z_0, z_1 \in \C.
\ee
Geometrically, each spinor defines a 3-vector $\vec{V}(z)$:
\be
 | z \rangle \langle z| =\frac{1}{2} (\langle z | z \rangle 1\! \!1    +\vec{V}(z)\cdot \vec{\sigma} ).
\label{3vector}
\ee
We will use the holomorphic representation of SU(2)  \cite{Livine:2011gp, Bargmann, Schwinger}. The representation space is the Bargmann-Fock space $L^2_{hol}(\C^2,d\mu)$  of holomorphic polynomials of a spinor, with Gaussian Hermitian inner product:
\be
  \bra f | g \ket = \int_{\C^2} \overline{f(z)} g(z) \rd\mu(z),
\ee
where $\rd\mu(z) = \pi^{-2} e^{-\bra z | z \ket}  \rd^{4}z$ and $\rd^{4}z$ is the Lebesgue measure on $\C^2$.

Holomorphic polynomials with different degrees of homogeneity are orthogonal with each other. Irreducible representations of spin $j$ are given by holomorphic functions with homogeneity degree of $2j$ in the $2j+1$ dimensional subspace: 
\be
L^2_{hol}(\C^2,d\mu)=\bigoplus_{j\in \N /2} V^j.
\ee
The $n$-valent intertwiners are a basis of $SU(2)$ invariant functions of n spinors. We denote the Hilbert space of $n$-valent intertwiners:
\be
  \cH_n \equiv \bigoplus_{j_i} \text{Inv}_{\text{SU}(2)}\left[V^{j_1} \otimes \cdots \otimes V^{j_n} \right] .
\ee
One way to construct elements in $\cH_n $ is by using the  Haar projector, which maps holomorphic polynomials of $n$ spinors to the $SU(2)$ invariant subspace $P :L^2(\C^2,\rd\mu)^{\otimes n} \rightarrow \cH_n$:
\be
  P(f)(w_i) = \int \prod_i \rd\mu(z_i) P(\check{z}_i;w_i) f(z_1, z_2, ... , z_n) = \int_{\text{SU(2)}} \rd g f(g w_1, g w_2, ... , g w_n),
\ee
the kernel of which is given by group averaging of the spinors with respect to the Haar measure over $SU(2)$ \cite{Freidel:2010tt, Freidel:2012ji} 
\be
 P(z_i;w_i) = \int_{\text{SU}(2)} \rd g \,e^{\sum_i [z_i|g|w_i\ket} = \sum_{J=0}^\infty \frac{\left( \sum_{i<j} [z_i|z_j\ket[w_i|w_j\ket\right)^J}{J!(J+1)!}.
\label{proj}
\ee
The second equality comes from performing the group integration \cite{Freidel:2012ji}. It is easy to check that $P(z_i;w_i) $ satisfies the projection property:
\be
 \int \prod_i \rd \mu(w_i) P(z_i;w_i) P(\check{w}_i;z'_i) = P(z_i;z'_i) .
\label{projection}
\ee
For the Riemannian 4d Spin Foam models, the gauge group is $Spin(4) = SU(2)_L \times SU(2)_R$, which is the double cover of $SO(4)$.  The partition function of $Spin(4)$ BF theory is a product of the two $SU(2)$ sectors. Classically, the 3-vectors defined by the spinors  [\ref{3vector}] from the two $SU(2)$ representations correspond to the selfdual and anti-selfdual components of the $B$ field \cite{Dupuis:2011fz}.  To obtain the model for 4-d gravity, we have to impose the simplicity constraints on the BF partition function. The holomorphic simplicity constraints, which were introduced in \cite{Dupuis:2011fz}, are essentially an isomorphism between the two spinor representation spaces of $SU(2)$.  The statement is that for any two links $i,j$ from the same intertwiner $a$, the holomorphic products of two spinors are proportional to each other:
\begin{equation}
[z^a_{iL} | z^a_{jL} \ket = \rho^2 [z^a_{iR} | z^a_{jR} \ket,
\label{hsc}
\end{equation}
where $\rho \in (0,1)$ is a function of the Immirzi parameter $\gamma$:
\begin{equation}
\rho^2 = \left\{ 
  \begin{array}{ll}
 (1-\gamma)/(1+\gamma), & \quad  |\gamma| < 1\\
 (\gamma-1)/(1+\gamma), & \quad  |\gamma| > 1.
  \end{array} \right.
\end{equation}
The difference with EPRL/FK models is that in this model the simplicity constraint is imposed on the $Spin(4)$ projector rather than the boundary state \cite{Banburski:2014cwa, Banburski:2015kmc}. The new \emph{constrained propagator} $P_\rho$ is defined as 
\be
\begin{split}
  P_\rho (z_i;w_i) &\equiv P(z_i;w_i) P(\rho z_i;\rho w_i) =\sum_{J} F_\rho(J) \frac{\left( \sum_{i<j} [z_i|z_j\ket[w_i|w_j\ket\right)^{J}}{J!(J+1)!} \label{cp}\\
& = \sum_{\{j_i\}}  F_\rho (J)  \int_{\text{SU}(2)} \rd g \prod^4_{i=1}\frac{ [z_i|g|w_i\ket^{2j_i}}{{2j_i}!} ,
\end{split}
\ee
in which the numerical factor in each $J = \sum_i j_i$ channel is actually the power series expansion of the hypergeometric function
\be
F_\rho(J):=  {}_2F_1(-J-1,-J;2;\rho^4) = \sum_{J'=0}^{J} \frac{J!(J+1)! \rho^{4J'}}{(J-J')!(J-J'+1)!J'!(J'+1)!}.
\ee
Here we can see that the constrained propagators are just BF projectors with non-trivial weight $ F_\rho (J) $. However, they no longer safisfy the projection property Eq.(\ref{projection}). One can see that from composing the propagator twice $P_\rho \circ P_\rho $:
\be
P_\rho \circ P_\rho   (z_i;w_i) = \sum_{J} \frac{F_\rho(J)^2}{(1+\rho^2)^{2J}} \frac{\left( \sum_{i<j} [z_i|z_j\ket[w_i|w_j\ket\right)^{J}}{ J!(J+1)!}.
\label{cp2}
\ee

Note that the imposition of simplicity constraints on all of the spinors also forces the measure of integration on $\C^2$ to change to
\be
\rd\mu_\rho(z) := \frac{(1+\rho^2)^2}{\pi^{2}}e^{-(1+\rho^2)\bra z|z\ket} \rd^2 z .
\ee
 The factor of $(1+\rho^2)^2$ is added for normalization. It insures that 
\be
\int \rd\mu_\rho(z) =1.
\label{normalization}
\ee

\begin{figure} [h]
\centering
\includegraphics[width=0.9 \textwidth]{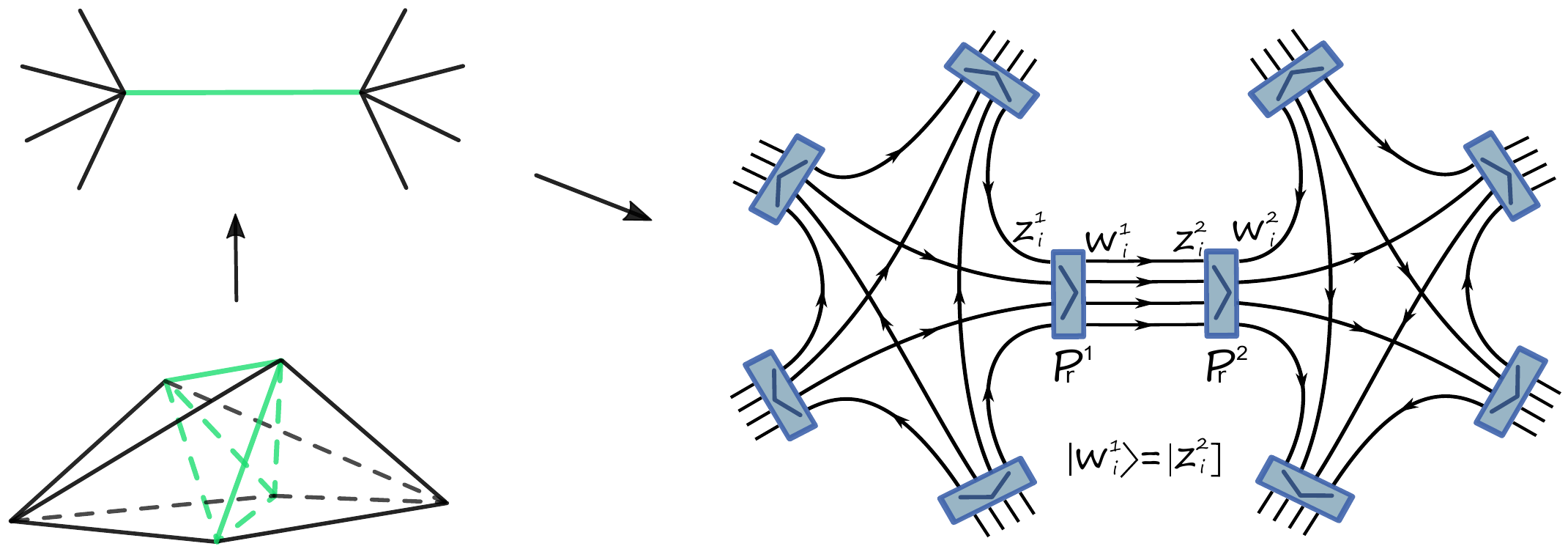}
\caption{The triangulation of two 4-simplices sharing one tetrahedron, the dual 2-complex and its cable diagram. The shared tetrahedron is dual to an edge propagator in the 2-complex. $P_\rho^1$ and $P_\rho^2$ belong to the same edge but two different 4-simplices. The spinors belonging on the same strand but belonging to different propagators are contracted according to the strand orientation. For example, spinors $w^1_i = \check{z}^2_i$.}
\label{2simplicescontract}
\end{figure}

 The partition function of the holomorphic spin foam model is defined on a 2-complex $\mathcal{G}  (V, E, F)$, which is dual to a simplicial decomposition of a 4-d manifold.  To make the geometrical relationship transparent, for each 2-complex we can also draw the corresponding cable diagram to label the degrees of freedom \cite{Perez:2012wv}. For an example of two 4-simplices sharing one tetrahedron, the dual 2-complex and its cable diagram, see Fig. \ref{2simplicescontract}.
In the cable diagram notation, a propagator is represented as

\be
  P_\rho (z_i;w_i) \equiv
\begin{tikzpicture}[baseline=0,scale=0.45]
  \node at (-4,1.5) {$[z_1|$}; \draw (-3,1.5) -- (-1,1.5);
  \node at (-4,0.5) {$[z_2|$}; \draw (-3,0.5) -- (-1,0.5);
  \node at (-4,-0.5) {$[z_3|$}; \draw (-3,-0.5) -- (-1,-0.5);
  \node at (-4,-1.5) {$[z_4|$}; \draw (-3,-1.5) -- (-1,-1.5);
  \draw (-1,-2) --(-1,2) -- (1,2) -- (1,-2) -- (-1,-2);
  \draw (1,1.5) -- (3,1.5); \node at (4,1.5) {$|w_1\ket$}; 
  \draw (1,0.5) -- (3,0.5); \node at (4,0.5) {$|w_2\ket$}; 
  \draw (1,-0.5) -- (3,-0.5); \node at (4,-0.5) {$|w_3\ket$}; 
  \draw (1,-1.5) -- (3,-1.5); \node at (4,-1.5) {$|w_4\ket$}; 
\end{tikzpicture},
\ee
where a strand represents a spinor, and a box represents group averaging with respect to the Haar measure over $SU(2)$, which is reduced from the projector of $Spin(4) = SU(2)_L \times SU(2)_R$ by simplicity constraints.

The faces $F$ in the dual 2-complex correspond to $d-2$  faces shared by tetrahedra. In the cable diagram, they correspond to loops formed by single strands.  The stucture of the partition function is essentially the contraction of the constrained propagators with non-trivial face weight:
\be
\mathcal{Z} (\mathcal{G} )=\prod_{f \in F}\sum_{j_f} \mathcal{A}_f(j_f) \int \left\{\prod_{all}\rd\mu_\rho(z)\rd\mu_\rho(w)\right\}\prod_{e\in\Delta^\ast}  P_{\rho} (z^e_i;w^e_i),
\label{Z}
\ee
where $\mathcal{A}_f(j_f)$ is the face weight, which is a function of spin $(2 j_f +1)^{\eta}$. With $\rho =0$ and $\eta =1$, the theory reduces to $SU(2) $ BF theory.  The partition function in principle can be defined for any type of discretization of the manifold: simplicial lattice, hypercubic lattice etc.  The rest of the paper is focusing on the discussion on simplicial lattice, but the result can be easily generalized to other cases. At the leading order, for a single simplex, the holomorphic spin foam model has the same semiclassical limit as EPRL/FK model \cite{Banburski:2015kmc}.

%


\subsection{Graph structure}
\label{Analyzing the graph structure}
In the 4-d spin foam model, the partition function is defined on a 2-complex  $\mathcal{G} (V, E, F)$, which is dual to a simplicial discretization of a 4-dimensional manifold.  The vertices $v \in V$ are dual to 4-simplices, edges $e \in E$ are dual to tetrahedra,  and the faces $F$  
of the 2-complex are dual to common triangles which are shared by 4-simplices.  Given a (connected) 2-complex $\mathcal{G} (V, E, F)$, there exist at least one \emph{spanning tree}  \cite{Kocay} $T_{\mathcal{G}} ( E_T) $ in the 1-skeleton of $ \mathcal{G} (V, E, F) $ and each of the trees contains $|V|-1$ branches $E_T$.

Now if we add one edge to the spanning tree, it will create a cycle. This type of cycle is called a \emph{fundamental cycle} in graph theory \cite{Kocay}.  Let us denote the set of fundamental cycles which are correlated with the spanning tree $T_{\mathcal{G}} (E_T)$ as $C_T$. For a given spanning tree, there is one to one correspondence between the edges not in the tree  and the fundamental cycles in $C_T$  \cite{Kocay}. The number of fundamental cycles $|C_T|$  is a tree-independent quantity:
\begin{equation}
|C_T| = |E\setminus E_T|= |E|-|V|+1
\label{CT}
\end{equation}

\begin{figure} [h]
\centering
\includegraphics[width=0.35 \textwidth]{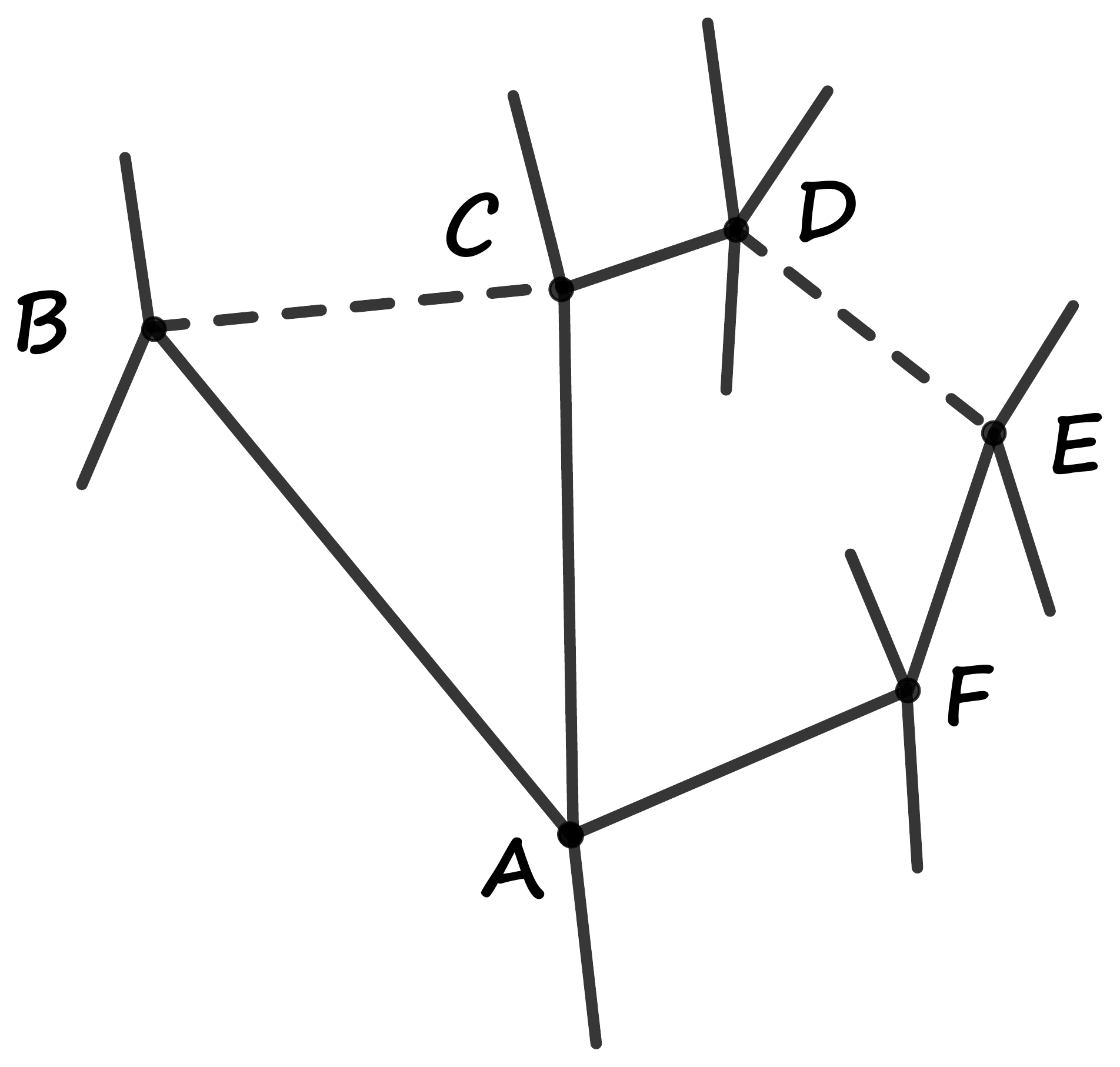}
\caption{The solid lines represent spanning tree $ T_{\mathcal{G}} = AB \cup AC \cup CD \cup EF \cup AF$. Edge $BC \in  E \setminus E_T$  corresponds to fundamental cycle  $ABC \in C_T$. Edge $DE$ corresponds to $ACDEF\in C_T$. }
\label{treecycle}
\end{figure}


For certain graphs, for example the diagrams of Pachner moves and melons, we can always choose a spanning tree such that all the corresponding fundamental cycles are also faces of the 2-complex (in cable diagram notation, this would be loops formed by the strands).  Let us call such type of spanning tree an \emph{optimal tree} and define  $\Omega_{\mathcal{G}}$ as a set of graphs which contain optimal spanning trees:
\begin{equation}
\Omega_{\mathcal{G}}\equiv\{  \mathcal{G} \ |\   \exists \ T_{\mathcal{G}} ( E_T) , \ s.t.\  C_{T} \subset F \}.
\label{ot}
\end{equation}

\begin{figure} [h]
\centering
\begin{minipage}{7.2 cm}
\includegraphics[width=0.9 \textwidth]{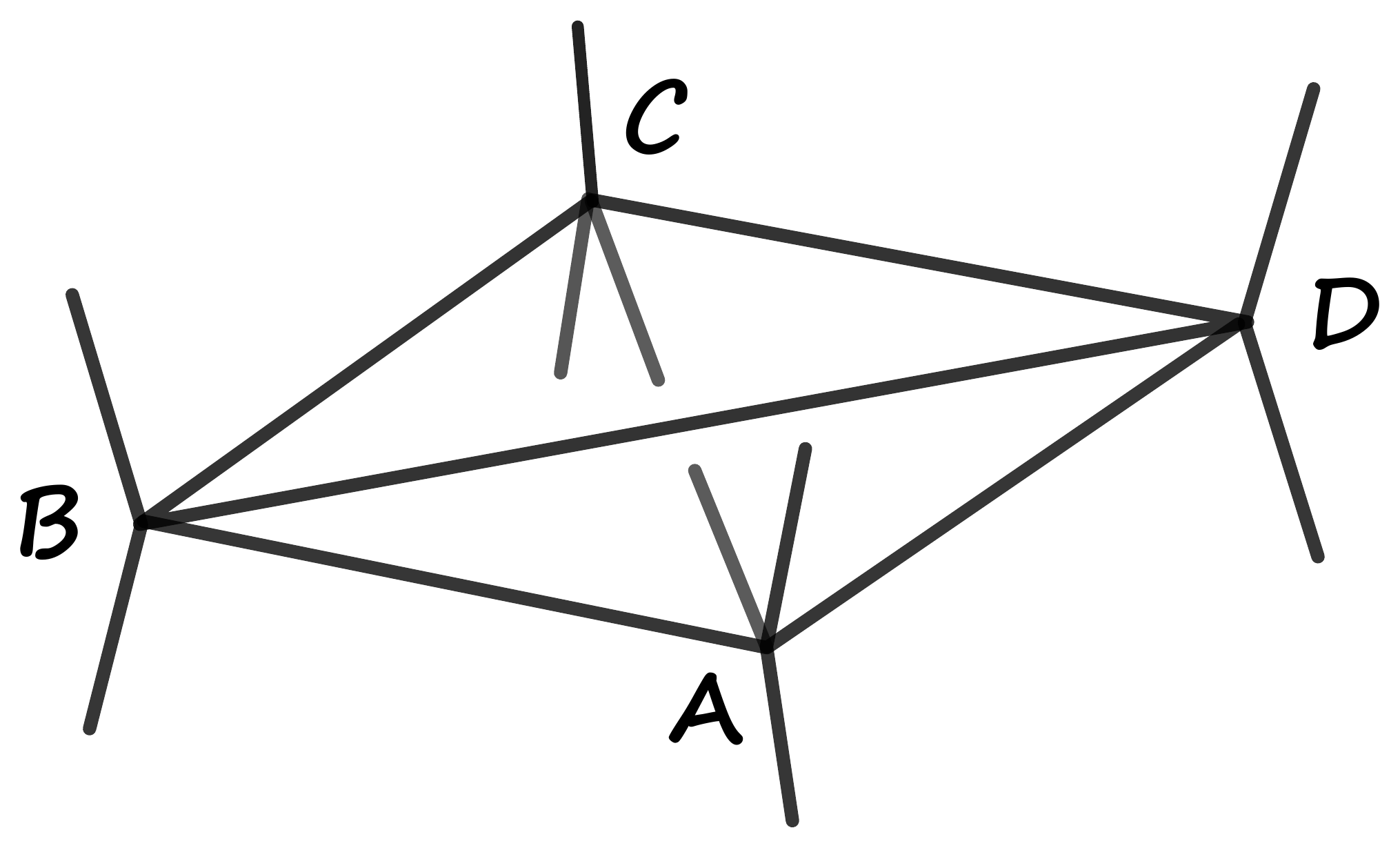}
\end{minipage}
\ \ \ \ \ \ \ \ \ \ 
\begin{minipage}{7.2 cm}
\includegraphics[width=0.9 \textwidth]{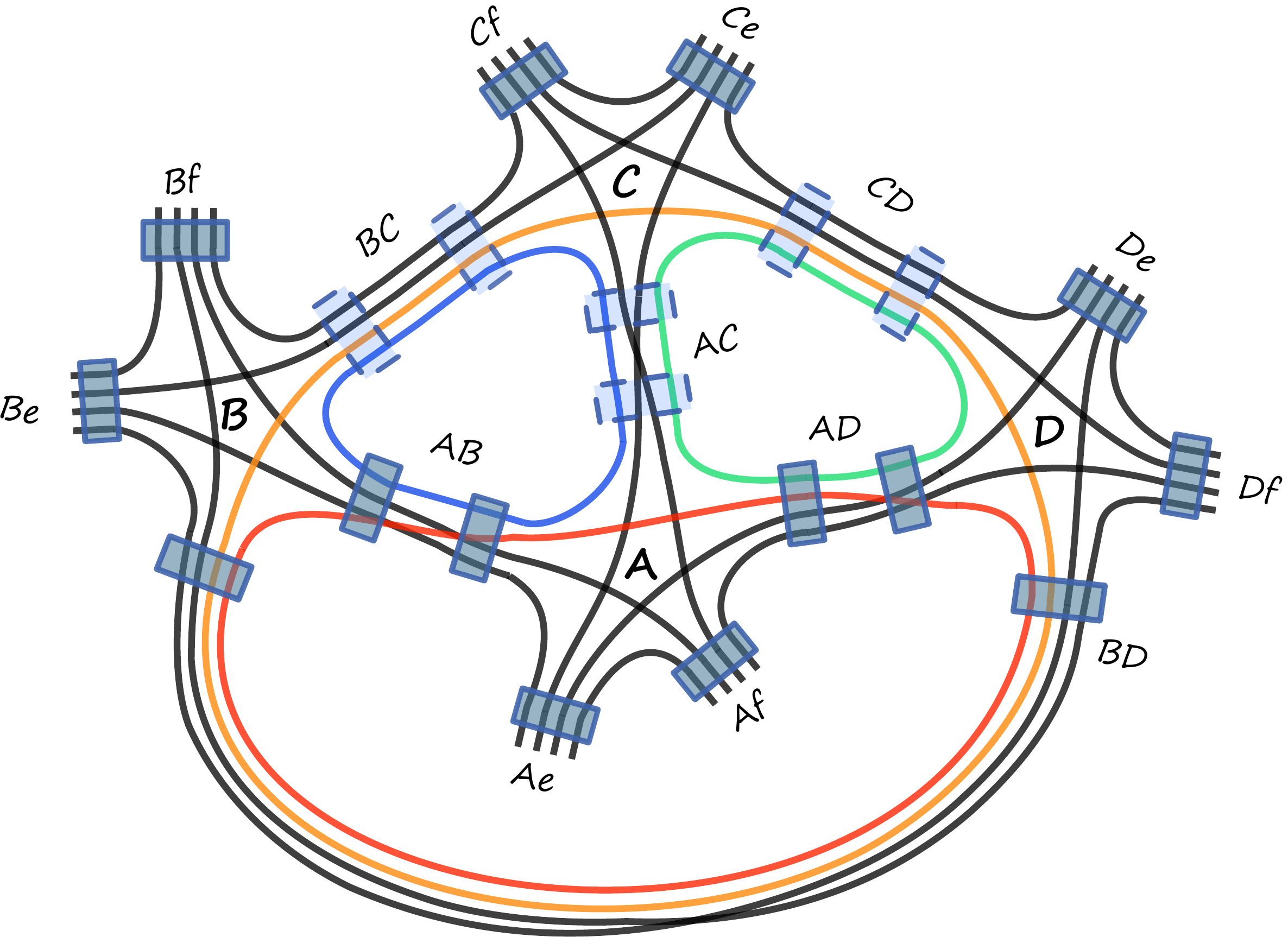}
\end{minipage}
\caption{ Take the graph of 4--2 Pachner move as an example: all the fundamental cycles of spanning tree $CA \cup CB \cup CD$ are faces of the 2-complex (single loops formed by strands in the right cable diagram): $ABC, ACD, BCD \in F$. However, for another choice of tree $CA \cup CB \cup AD$, one of the cycles $BCAD \notin F$. }
\label{optimal}
\end{figure}

In section \ref{Truncated bulk amplitude}, we will see that the existence of optimal spanning trees  in a graph is a very convenient property. We will show that the truncated bulk amplitude in such a graph can be written down just from reading out the combinatorics. 




\subsection{Partial gauge fixing}
\label{pgf}
In this section we briefly review a gauge fixing procedure which was first introduced and proved in \cite{Freidel:2002xb}. Given a (connected) 2-complex $\mathcal{G} (V, E, F)$ on which a partition function $Z(\mathcal{G})$ is defined, one can choose any spanning tree $T_{\mathcal{G}} ( E_T) \subset \mathcal{G}$ and gauge fix all the propagators on its edge $P_\rho^e, e\in E_T$ into
\begin{equation}
P_\rho (z_i;w_i) \rightarrow \id_\rho (\tilde{z}_i;\tilde{w}_i) \equiv \sum_{j_i}^\infty F_\rho (J)   \prod^4_{i=1}\frac{ [z_i|\one |w_i\ket^{2j_i}}{{2j_i}!},
\label{gf}
\end{equation}
leaving the partition function $Z(\mathcal{G})$ invariant.  Note that $J=\sum_i j_i$.

The invariance of the partition function can be proved through a systematic change of variables in the $SU(2)$ integrations of propagators. Note that for the convenience of notation later, we will always add a tilde on the spinors which belong to the partially gauge fixed propagator. Each spanning tree contains $|V|-1$ branches, thus we can gauge fix $|V|-1$ propagators based on Eq.[\ref{gf}] in a given 2-complex.

\subsection{The homogeneity map}
\label{homogeneity map}
As evaluating the partition function essentially amounts to integrations of power series of spinor polynomials, in \cite{Banburski:2014cwa} we introduced a useful tool - the homogeneity map - which allows us to perform the calculation in a more tractable and compact way. 

The basic idea of homogeneity map comes from the property that the holomorphic monomials of different degrees of homogeneity are orthogonal to each other. When we perform integration of a spinor, terms with different homogeneity do not mix:
\be
\int d\mu(w) [z_i|w \ket^{j} \ \langle w|z_2 ]^{l} =  j!\ [z_1|z_2]^{j} \  \delta _{j,l} .
\label{eg}
\ee
It would be convenient if instead of computing term by term in the power series, we could perform the integration in a compact form, with some book-keeping parameter to  keep track of each term with different homogeneity degree.  In this spirit, we define a general propagator $G_\tau$ in terms of an exponential form
\be
  G_\tau(z_i;w_i)  = e^{\tau \sum_{i<j}[z_i|z_j\ket [w_i|w_j\ket  } .
\ee
in which $\tau$  keeps track of the homogeneity of the polynomial in spinors.  In its series expansion, one can see that the following map transforms $G_\tau$ to a couple of constrained propagators (\ref{cp2})
\begin{align}
H_\rho: G_\tau \rightarrow P_\rho \circ P_\rho &  \qquad  \text{with} \qquad H_\rho:  \tau^J \rightarrow \frac{F_\rho(J)^2}{(1+\rho^2)^{2J} (J+1)!} \label{eqn_SC_map}
\end{align}
because the coefficients of each term in the power series  of  (\ref{cp2})  are functions of homogeneity degree $J$. The reason for considering $P_\rho \circ P_\rho $ comes from the partition function (\ref{Z}), in which every edge has a couple of propagators from two 4-simplices. Similarly we can define a trivial propagator $\one_{\tilde{\tau}}$ 
\be
\one_{\tilde{\tau}}=e^{\tilde{\tau} \sum_i [\tilde{z}_i|\tilde{w}_i\ket}.
\label{one}
\ee
The following homogeneity map recovers a couple of the gauge fixed propagators (\ref{gf}):
\be
 \tilde{H}_{\rho}: \one_{\tilde{\tau}} \rightarrow \one_\rho \circ \one_\rho  \qquad \text{with}\qquad    \tilde{H}_{\rho}: \tilde{\tau}^{J} \rightarrow \frac{F_\rho(J/2)^2 }{(1+\rho^2)^J} .
\ee
Using $\tau$ to keep track of homogeneity of each term, we first perform spinor integration with the exponential form such as $G_\tau$ and $\one_{\tilde{\tau}}$, then expand the result in terms of power series, and use the homogeneity map $H$ to restore the desired coefficients to get the final answer.


\subsection{Loop identity and truncation}
\label{loop identity and the truncation}
\begin{figure} [h]
\centering
\begin{minipage}{7.2 cm}
\includegraphics[width=1\textwidth]{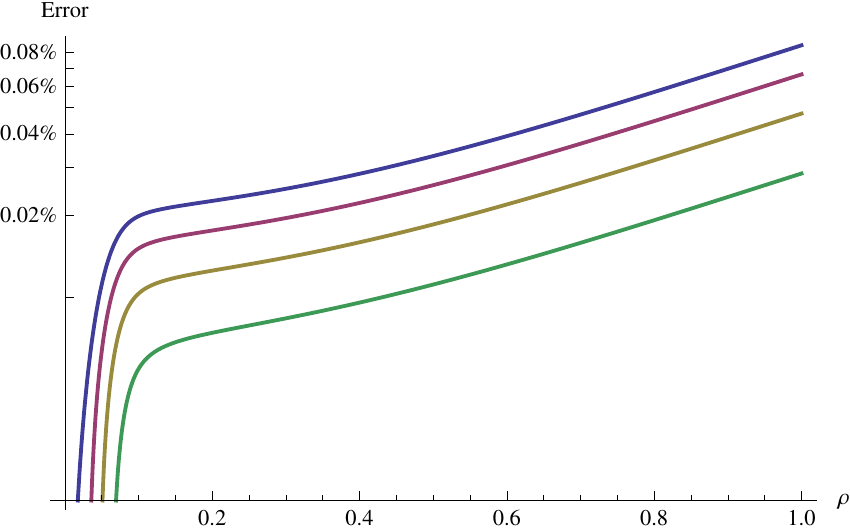}
\end{minipage}
\ \ \ \ \ \ \ \ \ \ 
\begin{minipage}{7.2 cm}
\includegraphics[width=1\textwidth]{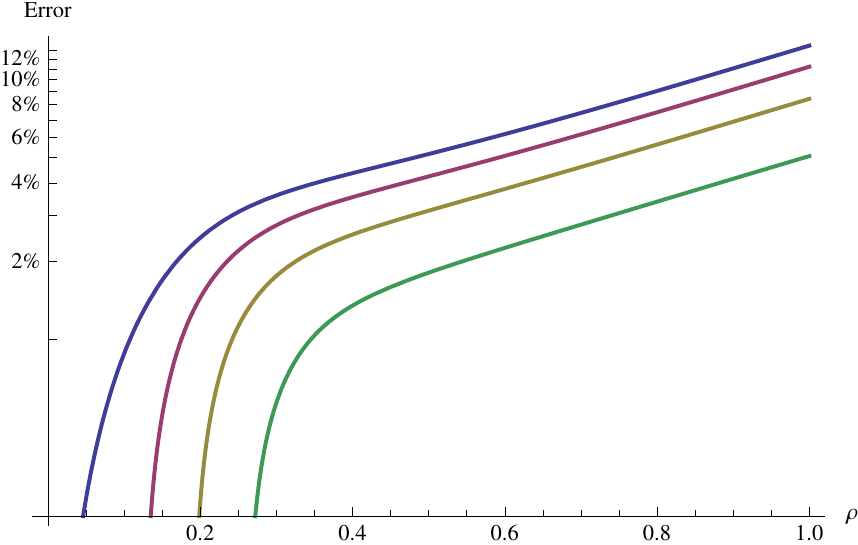}
\end{minipage}
\caption{Plots of the error from truncation for a single loop identity. The left plot is for large spin. The total spin on each propagator is 100. The right plot is for small spin, in which the total spin on each propagator is 5. Blue, red, yellow and green lines correspond to face weight $\eta$ equal to 1,2,3,4 respectively. The truncation is a better approximation for larger spins and larger face weights.
}
\label{truncation}
\end{figure}

To evaluate a partition function $Z(\mathcal{G})$ is to integrate out all the loops in its cable diagram. 
In \cite{Banburski:2014cwa}, we have calculated an identity for a partially gauge fixed loop with only one propagator not gauge fixed (see Appendix \ref{The Loop Identity}). It is crucial for evaluating Pachner moves in both 3-d and 4-d.   The special feature that differentiates BF theory and the spin foam model with simplicity constraints, is that integrating out loops results in mixing of strands:
\be
\raisebox{-8mm}{\includegraphics[keepaspectratio = true, scale = 0.33] {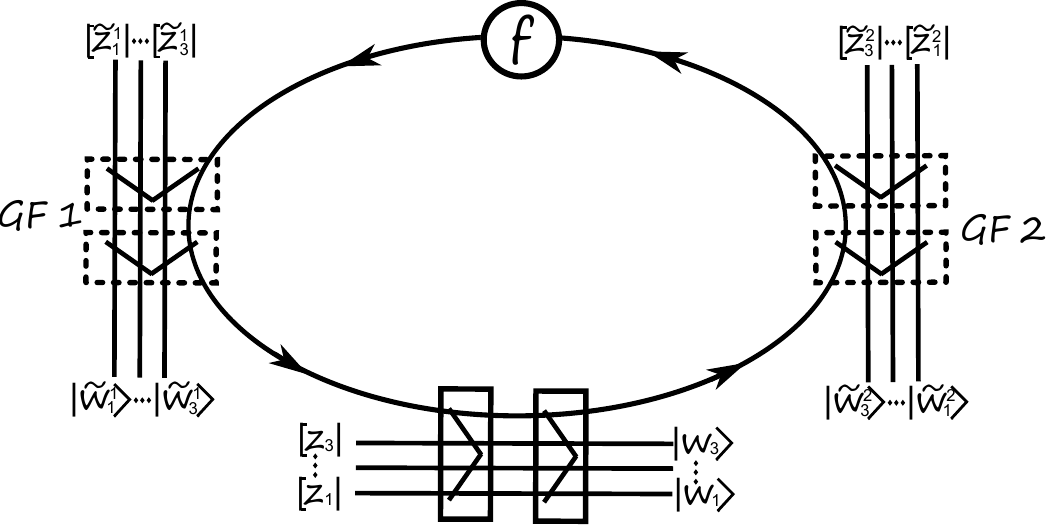}}=\sum_{A,B,J}N(J'=0) \!\!\!\!\raisebox{-8mm}{\includegraphics[keepaspectratio = true, scale = 0.3] {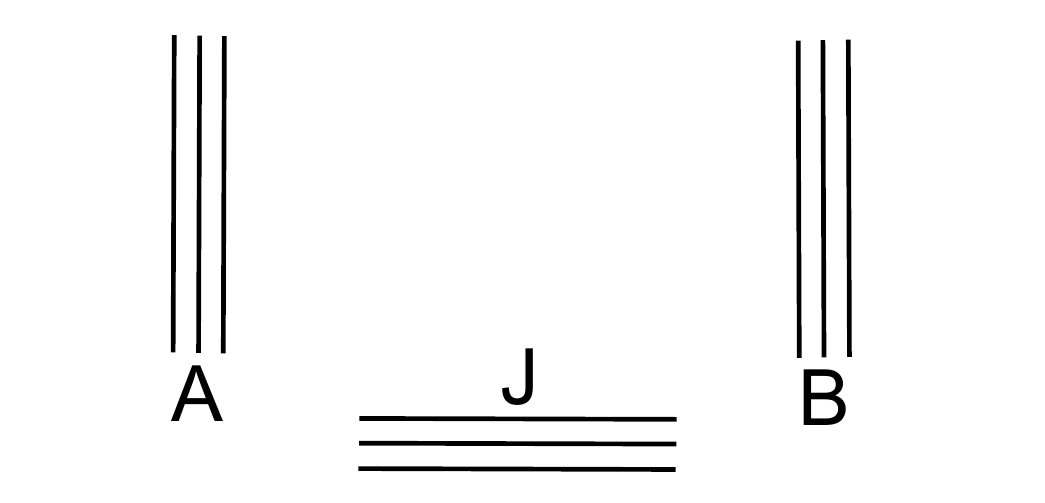}}\!\!\!\!\!\!+ \!\!\sum_{A,B,J,J'}N(J'\neq 0)\!\!\!\!\raisebox{-8mm}{\includegraphics[keepaspectratio = true, scale = 0.3] {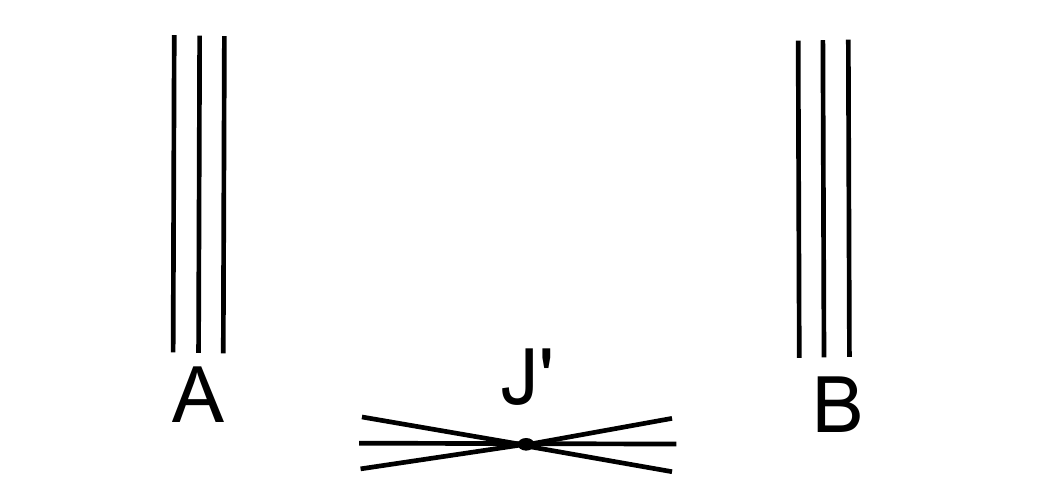}}\!\!\!\!\!\!.
\label{li}
\ee
In  \cite{Banburski:2014cwa}, it was shown that the mixing terms are sub-leading.  We introduced a natural truncation scheme, in which we  keep only the non-mixing term in Eq.(\ref{li}). The resulting amplitudes are structure preserving, and at the same time encode the non-local degrees of freedom as a non-local function of spins. The trunction scheme leads to small errors compared with the full amplitude, which is shown in Fig.\ref{truncation}

\begin{figure} [h]
\centering
\includegraphics[width=0.75 \textwidth]{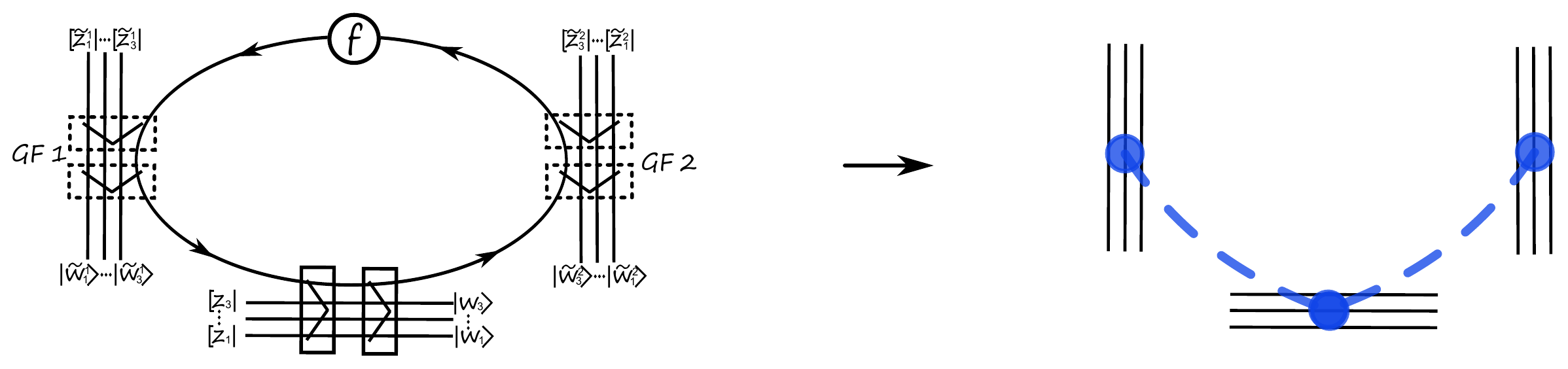}
\caption{We use blue dashed lines to represent the resulting non-local spin coupling from integrating out the loop.  }
\label{non-local}
\end{figure}

After truncation, the simplest and the most useful way of expressing the loop identity is in terms of an exponentiated form $L$ using the homogeneity map trick.  In general, for a partially gauge fixed loop with one original propagator and $\tilde{n}$ gauge fixed propagators labeled by $k=1...\tilde{n}$, the truncated loop identity is given by
\be\label{eq:exploopid}
L_\tau(z_i, w_i; \tilde{z}^k_i, \tilde{w}^k_i)=\exp  { \  \sum_{i=1}^{3}  \left( \sum_{k=1}^{\tilde{n}}\tilde{\tau}_k[\tilde{z}^k_i|\tilde{w}^k_i\ket  +T  [z_i|w_i\ket  \right)}.
\ee
We can see that this is essentially a product of $\tilde{n} +1$ trivial propagators (\ref{one}) with different book-keeping parameters  $\tilde{\tau}_k$ and $T$. The truncated loop identity can be recovered through applying the homogeneity map:
\begin{equation}
T^J \rightarrow \frac{J!  (J+1)^{\eta} \tau^J } {(1+\rho^2)^{ J\cdot (\tilde{n} +1)}  }  \  \left( \small{\prod}_k^{\tilde{n}}  \tilde{\tau}_{k}  \right)^J
\label{lihom}
\end{equation}
in which the propagators are tracked by
\begin{equation}
\tau^J  \rightarrow\frac{F_\rho (J)^2}{(1+\rho^2)^{2J} (1+J)!},\ \ \ \ 
 \tilde{\tau}^J_{k}   \rightarrow  \frac{F_\rho (J/2)^2}{(1+\rho^2)^J}.
\label{lihom2}
\end{equation}
From Eq.(\ref{lihom}), we can see that the homogeneity map  associated with the original propagator contains the information ($\tilde{\tau_k}$) from the gauge fixed propagators. Thus integrating out loops creates non-local spin couplings in 4-d spin foam amplitudes. We graphically represent this non-local coupling as in Fig.\ref{non-local}.

\newpage

\section{Truncated bulk amplitude}
\label{Truncated bulk amplitude}
In this section, we will study the \emph{bulk amplitude} - the evaluation of the partition function on a fully contracted 2-complex, or the evaluation on a connected 2-complex with zero boundary spins. This quantity is of interest because it isolates the degrees of freedom in the model which could lead to divergence. In spin foam amplitudes, the divergence comes from unbounded summations of spins, hence any finite boundary data is irrelevant for evaluating the degrees of divergence. 

In previous section \ref{Analyzing the graph structure} we have defined a certain set $\Omega_{\mathcal{G}}$ of 2-complexes, which contain optimal spanning trees.  We will first derive a formula of the truncated bulk amplitude for such type of graphs in this section. The formula will allow us to simply write down the truncated bulk amplitude by reading out the combinatorial properties of the graph. We then generalize the result to the arbitrary connected 2-complex. The results we obtain in this section will pave our way towards evaluating the divergence for arbitrary amplitude.

\subsection{Simple cases}
\label{truncated bulk partition function}
Now we will derive a formula for the truncated bulk amplitude of the 2-complexes in $\Omega_{\mathcal{G}}$. The graphs in $\Omega_{\mathcal{G}}$ are convenient to consider, because they possess optimal spanning trees such that all the corresponding fundamental cycles are faces of the 2-complex (loops formed by strands in the cable diagram notation).  The strategy to evaluate the amplitude of such a graph is as follows: first we gauge fix the propagators (\ref{gf}) along a chosen optimal tree $T_\mathcal{G} (E_T)$.  We denote the corresponding set of fundamental cycles as $C_{T}$. These cycles are loops formed by single strands in the cable diagram notation.  After gauge fixing, there are two types of propagators: the gauge-fixed propagators on the branches $E_T$ of the tree and the original propagators on the edges of the tree's complement $E \setminus E_T$.  The bijection between $C_{T}$ and $E \setminus E_T$ is justified by the fact that there is a unique original propagator per fundamental cycle, and on the other hand any original propagator belongs to a fundamental cycle.  We can apply truncated loop identity Eq.(\ref{lihom}) on elements of $C_{T}$.  The truncated loop identity annihilates all the fundamental cycles,  as well as $|V|-1$ number of gauge fixed propagators  and $|E| -|V|+1$ number of original propagators on the loops.  At this stage there are $|F|-|C_T|$ loops remaining with non-local functions coupling them through the homogeneity map (\ref{lihom}). Let us refer to these loops as \emph{residual loops} $L$.

Integrating out all the spinors in the residual loops and then applying homogeneity maps  give us the truncated bulk amplitude.  Without loss of generality, we use $T_\alpha$ to track the homogeneity of each original propagator $\alpha$ after applying the loop identity, use $l$ to label the residual loops, and use $n_l$ to denote the number of propagators in each of these residual loops. The structure of the exponentiated bulk amplitude $\mathcal{A} ^\tau _{bulk}$ is a product of $|F|-|C_T|$  spinor Gaussian integrations corresponding to the residual loops:

\begin{equation}
\begin{split}
\mathcal{A} ^\tau _{bulk} &= \prod_{l=1}^{|F|-|C_T|} \int d\mu_\rho(z_l) \ \exp \left[ \frac{ \tau'_l \cdot \prod^{n_l}_{\alpha=1} T_{\alpha} } {(1+\rho^2)^{n_l-1}}   \bra z_l|z_l\ket  \right]   \\
&=\prod_{l=1} ^{|F|-|C_T|} \sum_{j_l \in Z\!\!\!\!Z /2} (2j_l+1)  \left(  \frac{\tau'_l \cdot \prod^{n_l}_{\alpha=1} T_{\alpha}}{(1+\rho^2)^{n_l}}\right)^{2j_l},
\end{split}
\label{exponentiated amplitude}
\end{equation}
in which $\tau'_l$ keeps track of the face weight. The equality from the first to second line comes from series expansion of the integration result.  Applying the homogeneity maps (\ref{lihom}), (\ref{lihom2}) and ${\tau'_l}^{J} \rightarrow (J_l +1)^\eta $ to $\mathcal{A} ^\tau _{bulk}$ leads to the desired truncated bulk amplitude $\mathcal{A} _{bulk}$.


In each term of the series expansion, $T_\alpha$  is raised to a power of spins $\sum_{ l\in \Phi_\alpha} 2j_l$, which is a summation of spins from the residual loops coupled with the original propagator $\alpha$.  In other words, $\Phi_\alpha$ is the set of residual loops $L= F \setminus C_T$ which contain $\alpha \in E\setminus E_T$ as an edge. Hence the set $\Phi_\alpha$ associated with the original propagator $\alpha$ is defined as:
\be
\Phi_\alpha \equiv \{\ l \in L\ | \  \alpha \subset l  \ \}.
\ee
The  homogeneity map corresponding to the $T_\alpha$ comes from (\ref{lihom}) and (\ref{lihom2}). It is given by
\begin{equation}
T^{J} _\alpha \rightarrow \frac{ (J+1)^{\eta -1}  F_\rho (J)^2  } {(1+\rho^2)^{ J (3+\tilde{n}_\alpha)}}  \  \left( \small{\prod}_i^{\tilde{n}_\alpha}  \tilde{\tau}_{i}  \right)^J, \ \ \text{in which}\   J=\sum_{ l\in \Phi_\alpha} 2j_l ,
\end{equation}
in which $\tilde{\tau}$ tracks the homogeneity of gauge-fixed propagators.  Let us use index $k$ to label the gauge-fixed propagators.   In Eq.(\ref{exponentiated amplitude}), each $\tilde{\tau}_k$  is raised up to a power of spins  $\sum_{ l\in \Theta_k} 2j_l$. The set of summation contains spins from the residual loops which are connected with propagator $k\in E_T$ through a fundamental cycle $c_k\in C_T\ (k \subset c_k)$. Therefore we can define the set $\Theta_k$ associated with the gauge-fixed propagator $k$ as:
\be
\Theta_k \equiv \{  f \in  L \ | \ \exists\  c_k ,\  f \cap c_k \neq \emptyset \} .
\label{thetak}
\ee
The homogeneity map corresponding to the gauge-fixed propagator k in (\ref{exponentiated amplitude}) can be summarised as:
\begin{equation}
\tilde{\tau}^J_{k}   \rightarrow  \frac{F_\rho (J/2)^2}{(1+\rho^2)^J}, \ \ \text{in which}\   J=\sum_{ l\in \Theta_k} 2j_l .
\end{equation}

Before we write down the final result, let us however pause for a moment to introduce \emph{simplified diagrams} notation to make the structure simpler and more transparent. The simplified diagrams are reduced from the full cable diagrams by removing fundamental cycles, and only keep the combinatorial data we need for the final expression. They are very useful for us to write down the evaluation of $\mathcal{A}_{bulk}$ in a compact way.

\begin{figure} [h]
\centering
\includegraphics[width=1 \textwidth]{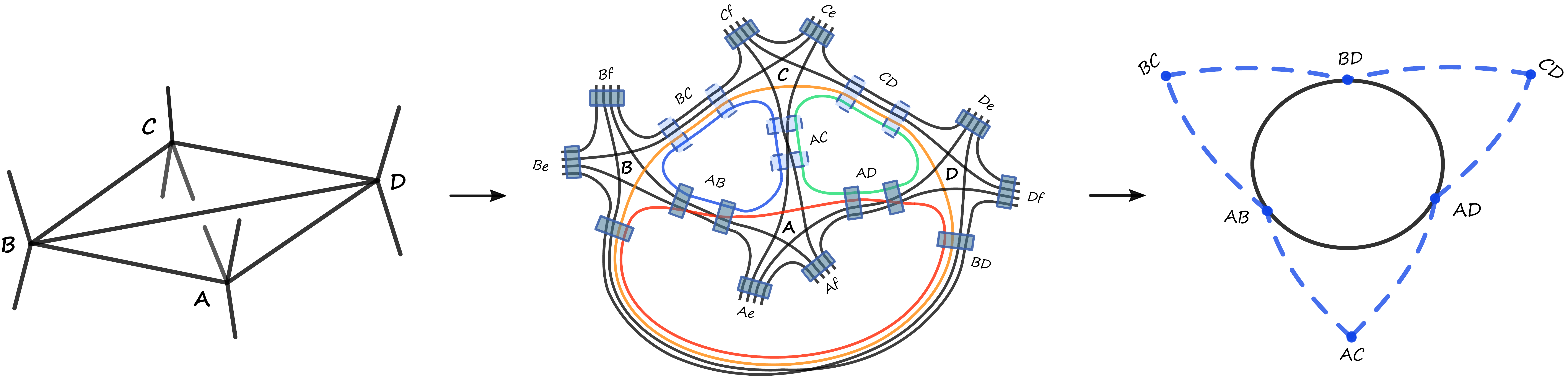}
\caption{We take 4--2 Pachner move as a simple example of showing how to reduce the cable diagram to the simplified diagram. Figures from left to right are the dual 2-complex, the cable diagram, and the simplifed diagram respectively.  With an optimal tree $CB \cup CA \cup CD$, the fundamental cycles are precisely the green, blue, yellow loops. 
Each gauge-fixed propagator is represented as a blue node in the simplified diagram, connected with two shared nodes on the residual loop ABD due to two loop identities. 
Thus evaluating the bulk amplitude for 4--2 Pachner move reduces to a single loop integration with non-local spin couplings, which are represented by the dashed lines.}
\label{42}
\end{figure}

\begin{figure} [h]
\centering
\includegraphics[width=1 \textwidth]{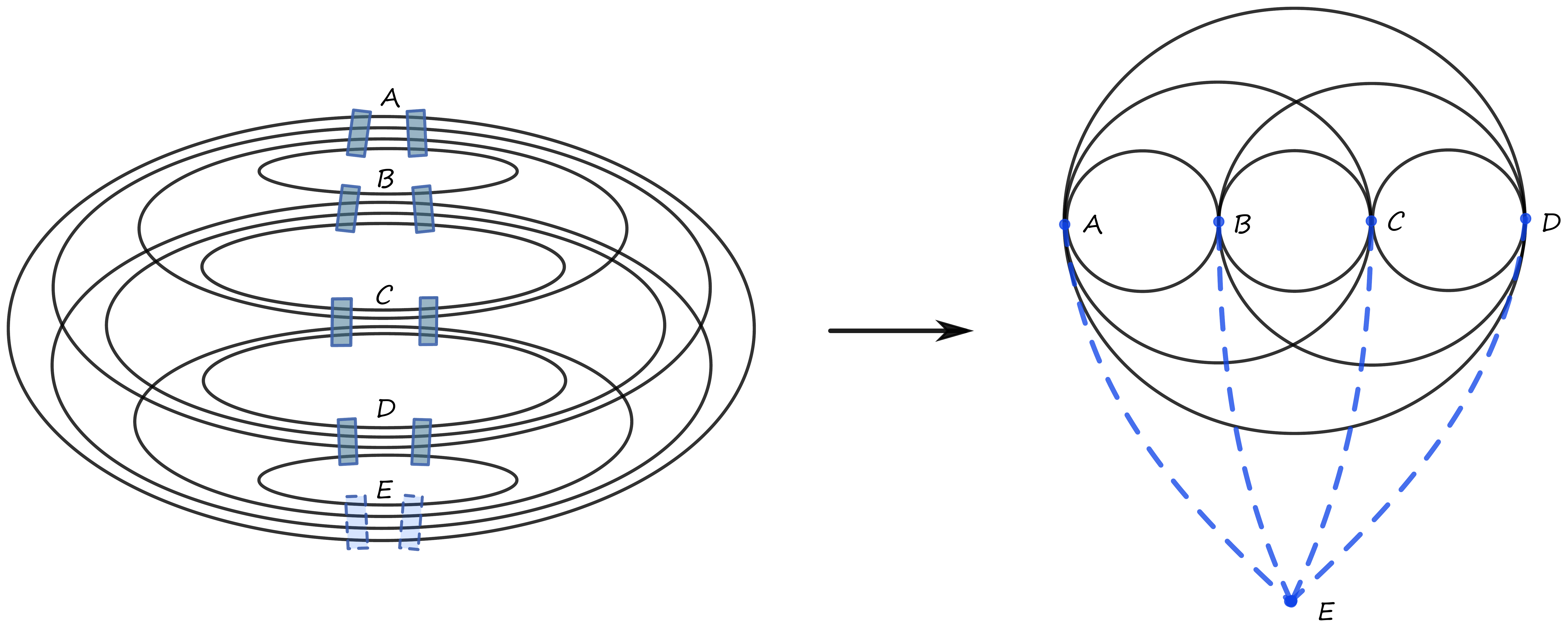}
\caption{Left figure: the cable diagram of a super melon (fully contracted two 4-simplices); right figure: the simplifed diagram. Optimal tree of this diagram only contains one branch and there are $|C_T|=4$ fundamental cycles corresponding to an optimal tree. We choose $E$ as the branch to be gauge fixed.  After applying the corresponding 4 loop identities there are $|F|-|C_T|=6$ residual loops. In the simplified diagram, the 6 residual loops coupled among the 4 shared nodes correspond to 4 original propagators.  The 4 blue dashed lines represent the non-local correlation produced by 4 loop identities. }
\label{melon}
\end{figure}

In a simplified diagram, the residual loops $L$ are represented by circles and propagators are represented by nodes.  The gauge-fixed ones, which are on the branches $E_T$ of the chosen spanning tree, have index k and are denoted by the nodes outside of the circles. 
The original propagators, which are on the edges of the spanning tree's complement $E \setminus E_T$,  have index $\alpha$ and are denoted by the nodes on the circles  in the simplified diagram.
We then use dashed lines (as in Fig.\ref{non-local}) connecting nodes to represent the non-local coupling among propagators resulting from the loop identities, as we have discussed in the last section. Each dashed lines can be specified by a pair of indices $\alpha k$. 
 We show two examples of reducing the cable diagram to the simplified diagram in Fig.\ref{42} and Fig.\ref{melon} .  When we have $|F|-|C_T| \geq 1$, there is more than one residual loop. In this case a few circles might share a same node, which corresponds to loops coupled at the same propagator, as in Fig.\ref{melon}. 

In the language of the simplified diagrams, the set $\Phi_\alpha$ contains the residual loops intersecting at node $\alpha$. $\Theta_k$ contains the residual loops which have connection through dashed lines $\alpha k$ with node k. Hence the definition of $\Theta_k$ (\ref{thetak}) can be written equivalently as:
\be
\Theta_k \equiv \{ \ f \in  L \ |\exists \alpha, \  \alpha k \cap f \neq \emptyset\  \}.
\ee

%
%


With all the above preparation, we can finally summarize the result of the truncated bulk amplitude in a very compact form:
\be
\!\!\!\!\!\! A_{bulk} \! = \!\!\!\!\sum_{\{j_l \in Z\!\!\!\!Z /2 \}} \prod_{l} \frac{(2j_l +1)^{\eta+1}}{(1+\rho^2)^{2j_l N_l}}\cdot \prod_{\alpha=1}^{|C_T|}  \left[F_{\rho}^2\ ( \sum_{ l\in \Phi_\alpha} 2j_l )\cdot (\sum_{l\in \Phi_\alpha } 2 j_l +1 )^{\eta -1} \right] \cdot \prod_{k=1}^{|V|-1} \left[ F_{\rho}^2\ ( \sum_{l\in\Theta_k } 2 j_l) \right],
\label{tba}
\ee
in which there are $|F|-|C_T|$ free summations of spins from the residual loops. The power of $(1+\rho^2)^{2j_l N_l}$ can be counted by the number of propagators $n_l$ in each residual loop, and the number of gauge fixed propagators $\tilde{n}_\beta$ in the fundamental cycle associated with the original propagator $\beta$:
\be
N_l \equiv 4 n_l + 2 \sum_{\beta=1} ^{n_l} \tilde{n}_\beta,
\label{N}
\ee
Actually from the simplifed diagram we can immediately read out the number $N_l$. This is because  $n_l$ counts the number of nodes on the residual loop $l$ and the summation $\sum_{\beta=1} ^{n_l} \tilde{n}_\beta$ counts the total number of dashed lines directly connected with the residual loop $l$.

The conveniece of Eq.(\ref{tba}) is such that given any 2-complex which belongs to the set $ \Omega_{\mathcal{G}}$, one can immediately write down the truncated bulk amplitude by simply counting the combinatorics in its simplified diagram without any calculation.  Let us look at a concrete example.

The left figure in Fig.\ref{melon} shows the cable diagram of two fully contracted 4-simplices, i.e. a super melon.  In this case, $|V|=2, |E|=5, |F|=10$. The optimal tree of the graph only contains one branch, which can be any of the five propagators.  By gauge fixing edge $E$ and removing the fundamental cycles through loop identities, we arrive at the corresponding simplified diagram, the right figure in  Fig.\ref{melon}.  In this case there are 4 fundamental cycles.  After applying 4 truncated loop identities, there are $|F|-|C_T|=6$ loops left that originally shared 4 propagators, which correspond to the shared nodes $A, B, C, D$ among loops in the simplified diagram. Each loop identity gives rise to a non-local correlation between the gauge-fixed propagator and the residual loops and they are represented by the blue dashed lines.

Now we can just read out the truncated bulk amplitude from the simplifed diagram: for each residual loop, there are two nodes on it, i.e. $n_l =2$ and two dashed lines directly connected with the loop, i.e. $ \sum_{\beta=1} ^{n_l} \tilde{n}_\beta =2$.  Thus from Eq.(\ref{N}), we have $N_l = 12$ for all the $l$. There are 6 independent summations of spins corresponding to the 6 residual loops l:

\be
\!\!\!\!\!\! A_{melon} = \!\!\!\!\! \sum_{\{ j_{AB},j_{AC}....j_{CD}\}} \prod_l \frac{(2j_l +1)^{\eta+1}}{(1+\rho^2)^{24 j_l }}\cdot  F_{\rho}^2\ ( \sum_{l\in \Theta_E} 2 j_l)  \cdot \prod_{\alpha}\left[ F_{\rho}^2\ ( \sum_{l\in \Phi_\alpha} 2j_l )\cdot (\sum_{l\in \Phi_\alpha} 2 j_l +1 )^{\eta -1} \right]
\label{bubbletba}
\ee
where $\alpha \in \{A, B, C, D\}$, the set $\Theta_E = \{AB, BC, AC, AD, BD, CD\},$
\be
\Phi_A=\{AB,AC,AD\},  \Phi_B=\{AB, BC, BD\},  \Phi_C=\{BC, CD, AC\}, \Phi_D=\{CD,BD,AD\}.
\ee
With this we can see the power of the simplified diagrams for 2-complexes in $\Omega _ {\mathcal{G}}$ -- the expression for the truncated amplitude depends only on the combinatorics of these diagrams. In the next section we will generalize the expression (\ref{tba}) for the amplitude to arbitrary connected 2-complexes, not only those with optimal spanning trees.


\subsection{The general structrue}
\label{The general structrue}
We have just shown that for a certain class of 2-complexes in $\Omega _ {\mathcal{G}}$, the truncated bulk amplitude can be read out through the combinatorics of their simplified diagrams. The convenience comes from the graph structure of elements of $\Omega _ {\mathcal{G}}$ -- the existence of the optimal spanning tree.  However, what if we do not choose the optimal spanning tree to fix the gauge?  Are the truncated degrees of freedom tree-dependent or not?  Would the truncated bulk amplitude  still be characterized by formula (\ref{tba})?   Moreover, in general for a graph that probably does not contain an optimal spanning tree, can we still express the truncated bulk amplitude in a simple form? 

If we choose a non-optimal tree, there exists at least one fundamental cycle which is not a loop formed by a single strand, but  a union of multiple loops coupled together in the cable diagram. In this case, we cannot directly apply loop identity to such a fundamental cycle even though it only contains one non-gauge-fixed propagator.  We need to generalize the loop identity in section \ref{loop identity and the truncation} to embrace such situations.

\begin{figure} [h]
\centering
\includegraphics[width=0.9 \textwidth]{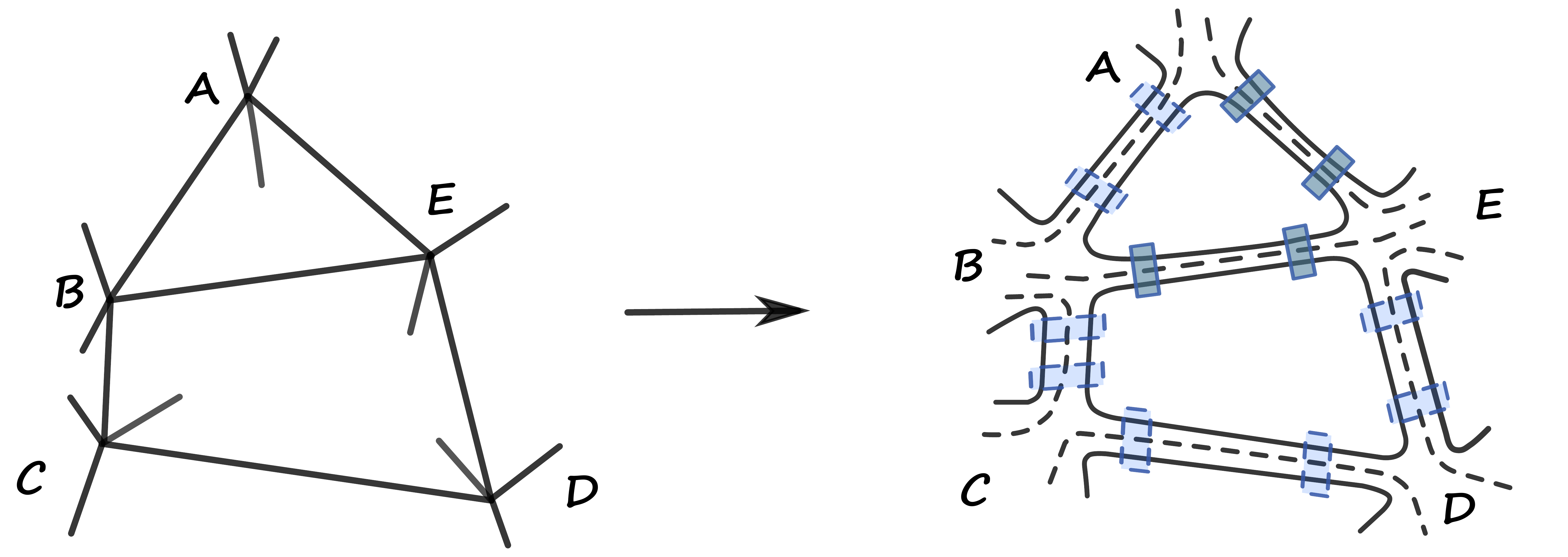}
\caption{With a choice of spanning tree $AB \cup BC \cup CD \cup DE$,  one of its fundamental cycle $ABCDE$ nests two loops $ABE\cup BCDE$.  We cannot directly apply loop identity for either $ABE$ or $ABCDE$. To integrate out all the loops in the cable diagram, they have to be annihilated in a specific order.}
\label{nl}
\end{figure}

Let us consider an example in Fig.\ref{nl}. In this case, if we choose a spanning tree as $AB \cup BC \cup CD \cup DE$, adding an edge $AE$ would create a fundamental cycle $ABCDE$, but it is not a face of the 2-complex. The cycle $ABCDE$ nests two loops $ABE\cup BCDE$ in the cable diagram.  To integrate out all the loops in the cable diagram in this case,  the loops need to be annihilated in specific order:  First we apply loop identity to $BCDE$ because it contains only one original propagator $BE$.  The truncation scheme neglects the mixing strands on edge $BE$,  thus after the truncation  we are able to apply loop identity again to $ABE$. In this way, we are evaluating a nesting of loop identities. The truncation for $ABE$ is performed after the truncation of loop $BCDE$, thus nesting of loop identities leads to truncation within truncation.

The homogeneity map of the nesting loop identity can be generalized from Eq.(\ref{lihom}).  In our example, for the loop $ABE$ in Fig.\ref{nl},  its exponentiated form is
\be\label{eq:gexploopid}
L_\tau=\exp  { \  \sum_{i=1}^{3}  \left( \tilde{\tau}_{AB}[\tilde{z}^{AB}_i|\tilde{w}^{AB}_i\ket +T_{BE}  [z^{BE}_i|w^{BE}_i\ket  +T_{AE}  [z^{AE}_i|w^{AE}_i\ket  \right)}.
\ee
This expression is of the same general structure as the loop identity (\ref{eq:exploopid})   -- it is a product of trivial propagators with different weights. The difference resides in the homogeniety map:

\begin{equation}
\mathcal{T}_{AE}^J \rightarrow \frac{J!  (J+1)^{\eta}} {(1+\rho^2)^{ 3 J}  }  \  \left(   \tilde{\tau}_{AB}  \tau_{AE}  T_{BE} \right)^J
\label{nlihom}
\end{equation}
in which the nested loop and the propagators are tracked by
\begin{equation}\label{map}
\begin{split}
 &T_{BE}^J \rightarrow \frac{J!  (J+1)^{\eta}} {(1+\rho^2)^{ 4 J}  }  \left(   \tilde{\tau}_{BC} \tilde{\tau}_{CD} \tilde{\tau}_{DE}    \tau_{BE} \right)^J, \\
\tau_{\alpha}^J  \rightarrow &\frac{F_\rho (J)^2}{(1+\rho^2)^{2J} (1+J)!},\ \ \ \ \ 
 \tilde{\tau}^J_{i}   \rightarrow  \frac{F_\rho (J/2)^2}{(1+\rho^2)^J}, \\
\end{split}
\end{equation}
where the indices $\alpha \in \{ AE, BE \}$ and $i  \in\{ AB, BC, CD, DE \}$. Compared with the loop identity (\ref{lihom}) we have discussed in Section \ref{loop identity and the truncation}, the only difference here for the loop $ABE$ is that the homogeniety map for the spinors on edge $BE$  is keeping track of a whole loop identity from loop $BCDE$. When we calculate nesting of loops, we need to apply loop identities in a specific order, and replace the homogeneity map of a gauge-fixed propagator in the second loop $ABE$ to the map which tracks the truncated loop identity in the first loop $BCDE$ . 

Since the generalization to the nesting of loop identities is straightforward, we can now evaluate the bulk amplitude without an optimal tree.  In a general graph, after gauge fixing along a spanning tree, there are $|E|-(|V|-1)$ original propagators to be integrated.  This number is still equal to $|C_T|$, which is the same as in the case of the class of graphs $\Omega_{\mathcal{G}}$. The only difference is that we need to consider nesting of loop identities, and annihilate loops in a  specific tree-dependent order.  After all the $|C_T|$ original propagators are annihilated through generalized loop identities, we have $|F|-|C_T|$ residual loops that have no group integrations to be performed anymore.  Performing gaussian integrals of spinors in those loops leads to $|F|-|C_T|$ number of independent summations of spins. 

At this stage, the result can be expanded in a power series. If we use $e$ to label edges, each book-keeping parameter $\tau_e$ is raised to the power of a summation of a few independent spins from the  residual loops: $\sum_{l\in\tilde{\Theta}_e } 2 j_l$.  After applying the homogeneity map to obtain the final result, the number of propagators $|E|$ in a graph corresponds to the number of squared hypergeometric functions $F^2 _{\rho} (\small{\sum}_{l\in\tilde{\Theta}_e } 2 j_l)$ in the amplitude.  
The set $\tilde{\Theta}_e$ is tree-dependent, and its elements can be straightforwardly obtained through the procedure outlined above.  Note that the spin $j_l$ of one residual loop appears in multiple propagators' $F^2_{\rho}$. This encodes the non-local feature of the result.

Similarly as in Eq.(\ref{tba}), in the general case we can also find out the power of $(1+\rho^2)$ explicitly, as it comes from two aspects: The first is $(1+\rho^2)^J$ as part of the normalization in each propagator (Eq.(\ref{map}) and Eq.(\ref{lihom})). The second is that when we integrate a spinor along a loop, each gaussian integral gives rise to $(1+\rho^2)^{-1}$ in the exponential:
\be
\int_{\C^2}\rd\mu_\rho (z) e^{\bra x|z\ket + \bra z|y\ket} = \sum_J  \frac{\bra x|y\ket^J}{J! (1+\rho^2)^{J} }  
\ee
After some algebra, one can check that the coefficient $N_l$ in the power of $(1+\rho^2)^{-2 j _l N_l}$ equals to twice of the number of $2 j_l$ appearing in the product of $F_{\rho} (\small{\sum}_{l\in\tilde{\Theta}_e } 2 j_l)$.

Thus finally, a general structure of the truncated bulk amplitude for an arbitrary connected 2-complex emerges and we summarize it as following:
\be
A_{bulk} = \!\!\!\!\! \sum_{\{j_l\} \in Z/2}  \prod_{e}\underbrace{ F_{\rho}^2\ ( \small{\sum}_{l\in\tilde{\Theta}_e } 2 j_l)} _{\# =|E|} \cdot \prod_f  [   (1+\rho^2)^{-2j_l N_l}  \cdot \underbrace{ (2j_l +1)^{\eta+1}}_{A: \#=|F|- |C_T|}  ]\cdot  \prod_{\alpha} \underbrace{ (\small{\sum}_{l\in \tilde{\Phi}_\alpha } 2 j_l +1 )^{\eta -1}} _{B: \#=|C_T|}
\label{tbageneral}
\ee

\begin{itemize}
\item There are $|F|-|C_T|$ free summations of spins.

\item The number of  $F_{\rho}^2\ ( \small{\sum}_{l\in \tilde {\Theta}_e } 2 j_l) $ is the number of propagators, i.e. $ \# F_{\rho}^2 =|E|$. The set $\tilde {\Theta}_e$ labels faces in the 2-complex and the details of the set $\tilde {\Theta}_e$ are tree-dependent.

\item The coefficient $N_l$ in the power of $(1+\rho^2)^{-2 j _l N_l}$ keeps track of  the occurence of each residual loop in the above sets  $\tilde {\Theta}_e$. In other words, it tracks the number of times $2 j _l$ appears in the product of $F_{\rho}$ and $\# |2j_l| = N_l/2$.

\item The number of $(2j_l+1)^{\eta+1}$ in the product equals to $|F|- |C_T|$.

\item The number of terms in the form of $(\sum_{l\in\tilde{\Phi}_\alpha} 2j_l+1)^{\eta-1}$  equals to $|C_T|$. $\tilde{\Phi}_\alpha$ is another set which labels the faces in the 2-complex. The details of the set $\tilde{\Phi}_\alpha$ are tree-dependent.
\end{itemize}

Now we can come back to the questions at the beginning of this section.  With different choice of gauge-fixing trees, the values of truncated bulk amplitude are different, which shows as different content in both of the sets $\tilde {\Theta}_e$ and $\tilde{\Phi}_\alpha$ in Eq.(\ref{tbageneral}). This is due to the truncation scheme.  In the loop identity, the mixing terms get truncated and those degrees of freedom depend on the chosen tree.  With nesting of loop identities as we have just discussed (for example Fig.\ref{nl}), the truncation within truncation usually leads to worse approximation than the truncation associated with an optimal spanning tree. In Appendix \ref{51move}, using the diagram of 5--1 Pachner move as an example, we evaluate the truncated bulk amplitude with different gauge fixing trees and compare the difference.  

We also recall that the truncation scheme is a better approximation for large spins. Hence if the bulk amplitude is convergent,  i.e. the most dominant degrees of freedom are given by small spin channel, the truncation is a worse approximation compared to the case when the bulk amplitude is divergent, i.e. large spin channels dominate the amplitude. Even though it is tree dependent, the truncated bulk amplitude of different graphs can be summarized with a same structure Eq.(\ref {tbageneral}). In the next section, we will derive a formula for the degrees of divergence based on this result Eq.(\ref {tbageneral}). We will see that the tree-dependent information (the details of the sets $\tilde{\Phi}_\alpha$ and $\tilde {\Theta}_m$) only contribute as finite factors in the large spin limit, and the most dominant degrees of freedom can be captured by a simple expression in terms of graph properties.


%
%
%


\section{Degree of divergence}
\label{Degree of divergence}
\subsection{The degree of divergence for arbitrary connected 2-complex}
In this section, we will show that the degree of divergence can be expressed only in terms of corresponding graph properties.   We will summarize the main result first, and then derive it in the later text.

For a connected 2-complex which is dual to a simplicial decomposition of 4-d manifold, the bulk degree of divergence is given by:
\begin{equation}
D(\mathcal{G})= \Lambda^{(\eta+2) |F| -3 |C_T| -3 |E| }, \ \ \ \ \  \text{when} \  |F| -|C_T| >0,
\label{result}
\end{equation}
where $\Lambda$ is a large spin cut-off. This expression can be rewritten, if we plug in the quantity of $|C_T|$ by Eq.(\ref{CT}), as
\begin{equation}
D(\mathcal{G}) = \Lambda^{(\eta +2) |F| - 6|E| +3| V| - 3}, \ \ \ \ \  \text{when} \  |F| - |E| + |V|-1 >0,
\label{result2}
\end{equation}
where $C_T$ is the number of fundamental cycles of an arbitrary chosen spanning tree in the graph,  $|F|$ is the number of faces in the 2-complex, $| V|$ is the number of vertices and $|E|$ is the number of edges in the bulk. 

In section \ref{Truncated bulk amplitude} we have seen that to get the bulk amplitude, we need to sum over independent spins in the $|L|=|F|-|C_T|$ residual loops. When $\Lambda \rightarrow \infty$ , integration is a good approximation of the summation:
\be 
\begin{split}
A_{bulk}(\Lambda)& \equiv \sum^{\Lambda/2}_{j_1\in \Z/2}...\sum^{\Lambda/2}_{j_{|L|}\in \Z/2}  f(2j_1, 2j_2,...2j_{|L|}) \\
&\approx \int^{\Lambda}_{\epsilon\cdot \Lambda}   f(j_1, j_2,...,j_{|L|}) \  dj_1...dj_{|L|} \\
& =  \Lambda^{|F|-|C_T|} \int^{1}_{\epsilon}  f(\Lambda J_1, \Lambda J_2,...,\Lambda J_{|L|})\  dJ_1...dJ_{|L|},
\end{split}
\label{g}
\ee
where we use an arbitrary small constant $\epsilon>0$ as small spin cut-off.  When the amplitude is finite, small spin regime dominates the amplitude. However, as we are seeking for the divergences, truncating small spin regime is irrelevant to the result, but it will help us to avoid some poles which come from $1/J^n$ in the asymptotic series expansion. 

It is easy to see that only when 
\be
|L|= |F| -|C_T| >0, \ \ i.e.|F| - |E| + |V|-1 >0
\ee
there are free summations in Eq.(\ref {tbageneral}), thus it is possible for an amplitude to diverge. In some cases, for example of the diagram in 3--3 Pachner move \cite{Banburski:2014cwa}, all the loops can be annihilated by the loop identity, thus there is no divergence associated with the diagram.

To derive (\ref{result}) let us look at the case $\rho =1$ first. In this case, our hypergeometric function is merely a rational function of factorials. In Appendix \ref{appenasy}, we have shown that it has asymptotics:
\be
{}_2F_1(-J -1, -J; 2 ; 1)= \frac{(2J+2)!}{(J+2)! \ (J+1)!} \sim \frac{4^{J+1} e }{\sqrt{\pi } \  J^{3/2}},  \ \ \ \ \ \ \  \text{as}\ \ J \rightarrow \infty.
\label{r1}
\ee
Note that this asymptotic formula is a good approximation even for small spins. We rewrite the bulk amplitude (\ref{tbageneral}) in the form of Eq.(\ref{g}) and then plug in the asymptotics (\ref{r1}).  For large spins, the asymptotic behavior of the bulk amplitude is given by
\begin{flalign*}
A_{bulk}(\Lambda) \sim &  \Lambda^{|F|-|C_T|}  \  \underbrace{\Lambda^{(|F|-|C_T|)\ (\eta+1)} }_{A}\  \underbrace{\Lambda ^{|C_T| (\eta-1)}}_{B}\  \left( \frac{ 4^{ \Lambda \sum_l  N_l}\  e^2 }{\pi  \  \lambda^3}  \right)^{|E|}   
 \   \prod_l  2^{-2\Lambda N_l}     
\times\\
& \times \int^{1}_{\epsilon}  \prod_{m}\left( \small{\sum}_{a\in\tilde{\Theta}_m } J_a \right)^{-3} \prod_l J_l ^{\eta+1}  \prod_{\alpha} \left( \small{\sum}_{b\in \Phi_\alpha } J_b  \right) ^{\eta -1} \  dJ_1...dJ_{|L|} .
\numberthis
\label{process}
\end{flalign*}
The integration in the second line is merely a finite number.  The underbraced parts A and B result from the corresponding parts in Eq.(\ref{tbageneral}).  Because the value of $N_l$ equals to twice of the number of $2j_l$ appearing in the product of hypergeometrical functions, we can see that the term of $4^{ \sum j_l N_l}$ which comes from the product of the asymptotics (\ref{r1}) exactly cancels the factor $\prod_l (1+\rho^2)^{-2j_l N_l} |_{\rho=1}$ in Eq.(\ref{tbageneral}).   Removing all the trivial constants from the above equation, we immediately arrive at 
\be
A_{bulk}(\Lambda) \sim \Lambda^{2 |F|-\eta |F| -3|C_T|-3 |E|} ,
\ee
which is exactly the result in Eq.(\ref{result}).

For $0<\rho <1$,  we have shown  in Appendix \ref{appenasy} that the hypergeometric function has the asymptotic expansion of
\be
 {}_2F_1(-J -1, -J; 2 ; \rho^4) \sim \frac{e^{\left(\frac{3}{2}+J\right) \zeta_\rho }\cdot \left(1-\rho ^4\right)^{\frac{3}{2}+J}}{2\sqrt{\pi }\  \rho ^3 \cdot J^{3/2} }, \ \ \text{for}\ \  0< \rho<1 \ \text{as}\ \  J \rightarrow \infty ,
\label{key}
\ee
where $\zeta_\rho \equiv \cosh^{-1}\left[ (1 + \rho^4)/(1 - \rho^4) \right]$.  Manifestly, this looks like a complicated function of $\rho$, but we importantly find that in the large $J$ limit, we have a tremendous simplification
\be
\lim_{J \rightarrow \infty} e^{\zeta_\rho J}\  (1-\rho^4)^J (1+\rho^2)^{-2J} =1 .
\ee
Thus similar to the case $\rho=1$,  the term of $ \prod_l e^{ j_l N_l \zeta_\rho }\  (1-\rho^4)^{j_l N_l}   $ which comes from the numerator of the product of the asymptotics (\ref{key}) cancels the factor $\prod_l (1+\rho^2)^{-2j_l N_l} $ in Eq.(\ref{tbageneral}) in the large spin limit.   With this cancellation, when we rewrite Eq.(\ref{tbageneral}) in the form of Eq.(\ref{key}), we get exactly the same expression (\ref{process}) for the bulk amplitude in the large spin limit. Hence we have just proved that for $0<\rho \leq 1$, the degree of divergence is characterized by (\ref{result}) and (\ref{result2}). Let us collect  a few simple examples of the degree of divergence in a table: 
\begin{figure} [h]
\centering
\includegraphics[width=0.9 \textwidth]{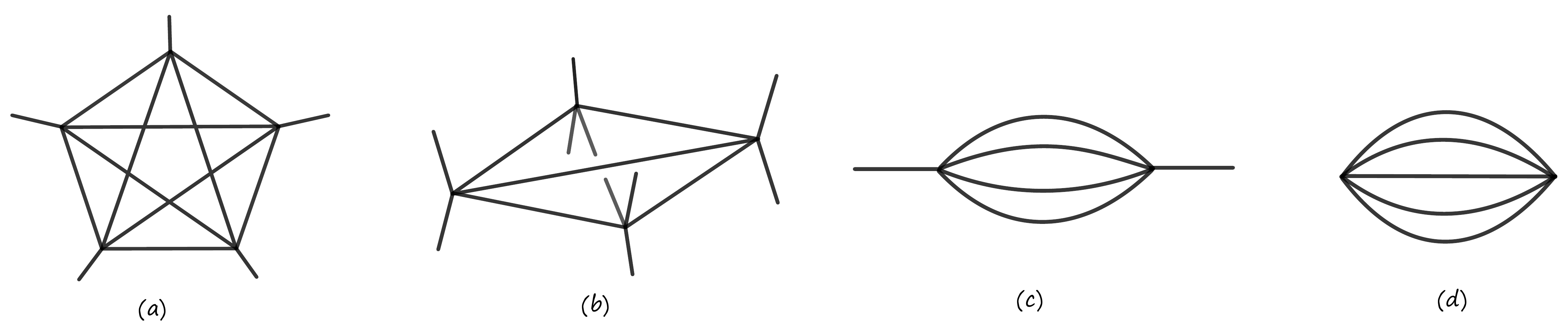}
\caption{A few simple examples we are considering in the following table.}
\label{4examples}
\end{figure}
\begin{center}
\begin{tabular}{ | m{12 em} || m{1cm}| m{1cm} | m{1cm} | m{10em} |} 
\hline
\ \  & \ $|V|$ & \ $|E|$ & \ $|F|$ & \ Degree of divergence  \\ 
\hline
5--1  move (a) & 5 & 10 & 10 &  $\ \ \ \Lambda^{10\eta-28}$\\ 
\hline
4--2  move (b) & 4 & 6 & 4 & $\ \ \ \Lambda^{4\eta-19}$   \\ 
\hline
Elementary melon (c)  & 2 & 4  & 6& $\ \ \ \Lambda^{6\eta-9}$\\ 
\hline
Fully contracted melon (d)  & 2 & 5  & 10&  $\ \ \ \Lambda^{10\eta-7}$ \\ 
\hline
\end{tabular}
\end{center}

When $\eta =3$, the 5--1 Pachner move has exactly $\Lambda^2$ divergence, which is the expected degree of divergence if the model has diffeomorphism invariance \cite{Baratin:2006gy, Dittrich:2011vz}.  It is important to note that for this value of $\eta$, the 4--2  move is finite. In fact, it does not become divergent until $\eta > 19/4$.

This spin foam model was expected to be less divergent compared with EPRL/FK, due to the difference of imposing the constraint on the propagator rather than the boundary spin network resulting in a more constrained model.  This is justified by the case of the elementary melon: the self-energy diagram starts to be divergent when $\eta > 1.5$, which is less divergent compared to previous results in both Riemannian and Lorenzian EPRL/FK model \cite{Krajewski:2010yq, Riello:2013bzw}. However, in the range of parameter $\eta$ when the melons are finite, the 5--1 Pachner Move is also finite, which makes the restoration of diffeomorphism symmetry an obstacle.


\subsection{Degree of divergence in terms of topological quantities}
\label{degree}
To gain some physical insight into the result we have obtained, it is useful to rewrite the degree of divergence in terms of topological and combinatorial quantities.  To simplify the discussion, we consider the case of compact 4-d manifolds, which are dual to fully contracted 2-complexes. 

If we use $N_i$ to represent the number of $i$-dimensional simplices in the triangulation, the d-dimensional Euler characteristic $\chi$ is defined as
\be
\chi =\sum_{i=0}^{d} (-1)^i N_i .
\label{chi}
\ee
In the 4-d simplical compact manifold, the link of every 2k-simplex ($k=1,2$) is an odd dimensional sphere \cite{Gabrielli:1997zy}. Recall that in d dimensions $S^d$ has Euler characteristic $1+(-1)^d$,  hence  
\be
 \sum_{i=2k-1}^{4}  (-1)^i  \binom{i+1}{2k-1} N_i =0 , \ \ \  k=1,2.
\label{2klink}
\ee
Recall that  the number of 4-simplices in the manifold is the number of vertices in the dual 2-complex $N_4=|V|$,  the number of tetrahedra is the number of edges  $N_3=|E|$, and the number of  triangles is the number of faces $N_2=|F|$. Eq.(\ref{2klink}) gives us the following two independent relations:
\begin{subequations}
\begin{align}
       2 |E|&=5 |V|, \\
2 N_1 +4|E| &=3|F| +5|V| .
\end{align}
\label{2}
\end{subequations}

Using Eq.(\ref{chi})  and Eq.(\ref{2})  we can  rewrite the degree of divergence (\ref{result}) in terms of the Euler characteristic:
\be
\text{In terms of $\chi$, $|F|$ and $|N_0|$}:\ \ D(\mathcal{G}) = \Lambda ^{(\eta-4) |F| + 3\left( 4 N_0 -4 \chi -1 \right)}.
\label{N0Fchi}
\ee
Unlike for 2-manifolds, in 4 dimensions we know that Euler characteristic is not enough to specify the topology. There is however another integer characterizing the topology: the \emph{degree} of a graph $\mathcal{G}$ \cite{Gurau:2010ba, Gurau:2011xq, Gurau:2011aq, Bonzom:2011zz, Gurau:2011xp}, which respects the formula 
\be
\frac{2}{(d-1)!}\omega_{d} (\mathcal{G}) = \frac{d(d-1)}{4} |V| + d - |F| .
\ee
The meaning of this quantity is related with a class of subgraphs: the \emph{jackets}.  A jacket of a cable diagram $\mathcal{G}$  is a subgraph which contains all the vertices $V$ and edges $E$ of $\mathcal{G}$, but only a subset of faces. The degree $\omega_{d} (\mathcal{G}) $ of a graph is the sum of the genera of its jackets. Hence the degree $\omega_{d} (\mathcal{G})  \geq 0$, while the equality is saturated with $\mathcal{G}$  dual to a sphere $S^d$. The reciprocal statement holds only when $d=2$ \cite{Gurau:2011xq}.  For reviews and discussion of this quantity see \cite{Gurau:2010ba, Gurau:2011xq, Gurau:2011aq, Bonzom:2011zz, Gurau:2011xp}.
\\
For a 4-d simplicial manifold, the degree of the skeleton of the dual 2-complex $\mathcal{G}$ and its number of vertices and faces are related by
\be
\omega_{4} (\mathcal{G})/3 = 3 |V| - |F| +4.
\label{degree4}
\ee
Thus in 4-d, from equations (\ref{chi}), (\ref{2klink}) and (\ref{degree4}), it follows that the degree of the graph and the Euler characteristic are related by 
\be
\omega_4(\mathcal{G})/3 = 3\chi-3N_0 +|F|/2 +4.
\ee
When we fix both the degree $\omega_{4d}(\mathcal{G})$ and the Euler characteristic $\chi$ of a graph, knowing one of the variables  $|V|, |F|$ and $N_0$ will fix the other two.

The degree of divergence formula (\ref{result})  can be rewritten in terms of the degree of the graph $\omega_4(\mathcal{G})$:
\be
\text{in terms of}\  \omega_{4d}(\mathcal{G}), |F|: \ \ 
D(\mathcal{G}) = \Lambda ^{(\eta-2)  |F| - 4 \omega_{4d}(\mathcal{G}) /3+13},
\label{omegaf}
\ee
\be
\text{in terms of}\  \omega_{4d}(\mathcal{G}), |V|:  \ 
D(\mathcal{G}) = \Lambda ^{ 3(\eta -2)  |V|  -  (2 + \eta)  \omega_{4d}(\mathcal{G}) /3   + 4 \eta+5}.
\ee
If we fix the degree $\omega_{4d}(\mathcal{G}) $ and vary $|V|, |F|$, then $\eta =2$ marks a transition of behaviors. With $\eta >2$, $D(\mathcal{G})$ monotonically increases with $|V|$ and the opposite happens when $\eta <2$. 
At $\eta =2$, the degree of divergence purely depends on  $\omega_{4d}(\mathcal{G})$:
\be
D(\mathcal{G})|_{\eta =2} = \Lambda ^{ -4 \omega_{4d}(\mathcal{G}) /3   + 13}.
\label{eta2}
\ee
From Eq.(\ref{degree4}) and also from the fact that the degree of a graph is the sum of the genera of its jackets, we know that $\omega_{4d}(\mathcal{G}) /3$ is a non-negative integer.  If we fix the number of vertices $|V|$ in the graph, when $\eta<-2$, $D(\mathcal{G})$ monotonically increases with  $\omega_{4d}(\mathcal{G})$.  However, the model is completely convergent with such face weight, which will not give the expected degree of divergence for 5--1 move.  In the region $\eta > -2$, $D(\mathcal{G})$ monotonically decreases with  $\omega_{4d}(\mathcal{G})$ and reaches its maximum with $\omega_{4d}(\mathcal{G}) = 0$. This means that if one uniformly sums over all the possible graphs, the dominant contributions to the partition function are the graphs which are dual to the simplicial manifolds with spherical topology and at the same time with degree $\omega_{4d}(\mathcal{G}) = 0$.  In \cite{Gurau:2011xq} it was shown that this type of graphs with $\omega_{4d}(\mathcal{G}) = 0$ are melonic.  It is a class of graphs with maximal $|F|$ at fixed $|V|$, and their elementary subgraph is composed by a couple of simplices glued along all but one of their faces (see Fig.\ref{4examples} c).  They are the leading order contribution in the large $N$ limit of colored tensor models \cite{Gurau:2011xq}, and have been long suspected to be the most divergent configuration in spin foams \cite{Krajewski:2010yq, Riello:2013bzw}.

%


\subsection{Physical implications}
First let us compare our result with the colored tensor models, in which the dominant graphs and continuum limit have been studied in depth. We will briefly review their results first. In the case of the independent identically distributed model and the Boulatov Ooguri model (\cite{Gurau:2011xq}, \cite{Gurau:2011xp}), the amplitudes associated with a graph $\mathcal{G}$ are given by
\be
A^{i.i.d.} (\mathcal{G}) = (\lambda  \bar{\lambda})^{|V|/2} N^{d -\frac{2}{(d-1)!} \omega(\mathcal{G}) }, \ \ \ A^{B.O.}  (\mathcal{G})  = (\lambda  \bar{\lambda})^{|V|/2} N^{d-1 -\frac{2(d-2)}{d!} \omega(\mathcal{G}) },
\ee
where $N$ is a large parameter indicating the tensor size, $\lambda $ and $\bar{\lambda}$ are coupling constants. Note that in these models the coupling constant has been rescaled by a power of $N$, so that the amplitude is not increasingly divergent or suppressed by higher number of vertices, and the amplitude of a graph depends solely on its degree \cite{Gurau:2011xp}.  For both of the i.i.d and BO models, it has been shown that the leading order contribution of the $1/N$ expansion \cite{Gurau:2010ba,Gurau:2011xq, Gurau:2011aq, Gurau:2011xp,Bonzom:2011zz,Gurau:2013cbh} is governed by melonic graphs $(\omega(\mathcal{G})=0 $) \cite{Gurau:2011xq}.   It was further shown in \cite{Gurau:2013cbh} that the melonic dominance leads to branched polymers phase in the continuum limit. 

In terms of the dominant graphs, the spin foam model we are studying has the same behavior with the colored tensor models when the face weight $\eta = 2$, as the degree of divergence (\ref{eta2}) solely depends on a negtive power of $\omega_4(\mathcal{G})$. If one sums over all the diagrams with equal weight as in the colored tensor model, then we predict that the spin foam model has a branched polymers phase in the continuum limit at $\eta=2$. 

When the face weight $\eta\neq 2$ however, the degree of divergence has a non-trivial dependence on  $|V|$.  When $-2<\eta<2$, the amplitude is increasingly suppressed with higer number of vertices $|V|$, which indicates that the most divergent diagram is a single super melon  When $\eta>2$, the amplitude is increasingly divergent with higher number of vertices $|V|$, which indicates that the coupling constant should be rescaled through renormalization.  This region is of physical interest because it contains the range of parameter in which the 5--1 move is divergent while 4--2 is convergent.  However, since the dominant diagrams are melonic and they are geometrically degenerate, one might worry that the model is peaked on the configurations which do not describe smooth 4d geometry, if there is no restriction on the set of allowed diagrams.

To address this concern, let us have a look at  3-d gravity first. The degree of divergence for Ponzano-Regge model (3-d $SU(2)$ BF theory) is captured by
\be
D_{SU(2) BF}(\mathcal{G}) = \Lambda ^{3|F| -3|E| +3|V|-3}.
\label{3d}
\ee
The face weight in the model is chosen to be $(2j+1)$, because it is the only choice which preserves topological invariance. There are different ways of deriving (\ref{3d}). In the approach presented in \cite{Banburski:2014cwa} and this paper, we can see that  (\ref{3d}) arises from the following simple derivation: each residual loop contributes to the degree of divergence a factor of $\delta_{SU(2)} (\one) \sim \Lambda^3$, and in a conneted diagram the number of residual loops is given by $|L| =|F| -|C_T| = |F| -|E| +|V|-1$.

When we discretize 3-d gravity, the residual action of the diffeomorphism group acts at the vertices of the triangulation of a 3-d manifold as vertex translation symmetry. The 4--1 Pachner move has exactly degree of divergence $\Lambda^3$, which corresponds to the translation symmetry of placing the free vertex anywhere in the triangulation. A proper Faddeev-Popov gauge fixing procedure divides the amplitude by this divergence \cite{Freidel:2002dw}, thus leaving the model invariant under 4--1 move. 
\begin{center}
\begin{tabular}{ | m{11em} || m{1cm}| m{1cm} | m{1cm} | m{10em} |} 
\hline
\ \  & \ $|V|$ & \ $|E|$ & \ $|F|$ & \ Degree of divergence  \\ 
\hline
4--1  Pachner move & 4 & 6 & 4 &  $\ \ \ \Lambda^{3}$\\ 
\hline
A fully contracted melon & 2 & 4 & 6 & $\ \ \ \Lambda^{9}$   \\ 
\hline
\end{tabular}
\end{center}
Now let us rewrite the degree of divergence in terms of topological invariants and the graph degree $\omega_{3d}$. For the simplicial decomposition of 3-d compact manifold, the Euler characteristic is
\be
\chi = N_0 -|F| +|E|-|V|
\ee
and the degree of the graph is given by
\be
2 \omega_{3d} (\mathcal{G}) =3 |V| -2|F| +6.
\ee
Together with the relation $2 |E| = 4 |V|$ for a fully contracted 2-complex, we find that the degree of divergence can be equivalently expressed as
\be
D_{SU(2) BF}(\mathcal{G}) =  \Lambda ^{3 N_0 -3 \chi -3 } =\Lambda ^{3|V|/2 - 3 \omega_{3d} (\mathcal{G})  +6 } 
\label{3ddivergence}
\ee
From expression in terms of $N_0$ and $\chi$, we can see that the divergence is concentrated on the number of free vertices $N_0$ in the triangulation. From the expression in terms of the degree, we can see that without properly gauge fixing the vertex translation symmetry, the most divergent graphs also have degree $\omega_{3d} (\mathcal{G}) =0$ and hence are melonic.  

The gauge fixing procedure introduced in \cite{Freidel:2002dw} fixes the spins along a maximum tree of bones in the triangulation to be zero.  In the approach presented this paper, this maximum tree precisely corresponds to the residual loops in the simplified diagrams. Setting the spins in those residual loops to  zero simply removes all the divergence.  Thus the gauge fixing procedure ensures that all the graphs have finite amplitude.  Without properly gauge fixing the diffeomorphism symmetry and removing the divergence, one should not rush into the conclusion that the Ponzano-Regge model is peaked on melonic graphs. 

The continuum limit of 3-d gravity is fully described by the discrete model, which allows us to identify the gauge symmetry and leaves the model finite. However the question is very non-trivial in 4-d, since the diffeomorphism symmetry is only expected to be recovered in the continuum limit through renormalization. Nevertheless, considering that the discretized 4-d classical Regge action indeed has the vertex translation symmetry, in the quantum model we might be able to identify some residual gauge symmetry with non-compact gauge orbits, which could be the origin of the divergence in the 5--1 Pachner Move.  As we have just shown, with face weight $\eta=3$ the model has the expected degree of divergence if the diffeomorphism symmetry is recovered.  Hence the future research should clarify  whether the model indeed contains some residual gauge symmetry which results in the divergence. Properly identifying and fixing this residual gauge symmetry might completely remove the divergence and change the behavior of melonic dominance, as is the case in 3-d. 

The implications of the result in this paper are different for the two distinct approaches towards continuum limit: summing over all the possible diagrams \cite{Rovelli:2014ssa, Freidel:2005qe,Oriti:2006se}, or refining the partition function on a fixed lattice (a la Dittrich \cite{Dittrich:2012jq, Bahr:2014qza, Dittrich:2014ala}). 

For the summation approach, the melonic dominance might be resolved by gauge fixing, as we have just discussed. Another way is by putting a restriction on the space of diagrams which are to be summed over, and discarding the geometrically degenerate cases.  Such an approach will result in a different model than the current group field theories. It also requires finding a complete classification of 2-complexes which are dual to non-degenerate geometries and is an interesting mathematical question by itself.  A related question is whether the fully contracted diagrams (in the context of group field theory) can be interpreted similarly as in quantum field theory, where the sum of all the vacuum bubbles is a normalization factor for the physical correlation functions.

For the refining approach, there is no concern of the melonic dominance and one can focus on possible phase transitions which have already been indicated by the distinct behaviors of the model in different ranges of $\eta$. There indeed exists value of $\eta$ such that 5--1 move has $\Lambda^2$ divergence while 4--2 move is finite.  This is a promising sign of recovering diffeomorphism symmetry in the continuum limit.

\section{Conclusions}
In this paper, we have studied the evaluation of a 4-d spin foam partition function on an arbitrary 2-complex and its degree of divergence, with recently developed methods -- the homogeneity map, loop identity and its truncation. For certain class of graphs, in which there exist optimal spanning trees, we derived a formula to capture the dominant degrees of freedom in the partition function.  Using the formula, one can simply read out the evaluation of truncated bulk amplitude through combinatorial properties of a graph.  

We then generalized the result to arbitrary graphs and studied the structure of truncated bulk partition function. With the gauge fixing choice along an optimal spanning tree, the error of truncation is minimized. For a generic choice of spanning trees, we need to evaluate nesting of loop identities which leads to truncation within truncation.  We then showed that even though the truncated degrees of freedom depend on the choice of gauge-fixing tree, the dominant degrees of freedom are tree-independent and can be captured by a simple expression. Using the asymptotic formula of the hypergeometrical functions, we extracted a simple formula for the exact degree of divergence for arbitrary 2-complexes, in which the variables are the number of vertices $|V|$, number of faces $|F|$ and the number of edges $|E|$.  The dependence on Immirzi parameter has dropped out in the asymptotic analysis, which agrees with the previous results in the EPRL model \cite{Riello:2013bzw,Perini:2008pd,Krajewski:2010yq}. The only parameter in the degree of divergence formula is the power of the face weight $\eta$. When the face weight is $(2j+1)^3$, i.e. $\eta =3$, the 5--1 Pachner move has $\Lambda^2$ divergence, which is the expected degree of divergence if diffeomorphism symmetry is recovered.

To gain some physical insight, we expressed the degree of divergence formula in terms of Euler chracteristic $\chi$ and the degree of graph  $\omega (\mathcal {G})$.  We found that for the face weight $\eta > -2$, the most divergent graphs are the ones that have spherical topology and degree $\omega_{4d}(\mathcal {G}) =0$. It has been shown in \cite{Gurau:2011xq} that this type of  graphs are melonic, which is a class of graphs with maximal number of faces at fixed number of vertices.  When the face weight $\eta = 2$, the degree of divergence (\ref{eta2}) solely depends on a negtive power of $\omega_{4d}(\mathcal{G})$.  If one sums over all the diagrams with equal weight as in the colored tensor models, then we predict that the spin foam model has a branched polymers phase in the continuum limit at $\eta=2$.  When $\eta<2$, the amplitude is increasingly suppressed with higer number of vertices $|V|$, which indicates that the most divergent diagram is a single super melon ($\eta >-2$).   In the region of parameter of physical interest (the 5--1 move is divergent while 4--2 is convergent), the amplitude is increasingly divergent with higher number of vertices $|V|$.  We might need to introduce a new coupling constant for the vertex, and it should be rescaled through renormalization. If we do not put restriction on the allowed diagrams, one might be concerned that the model is peaked on geometrically degenerate melonic configurations.

If one takes the point of view that the continuum limit of Spin Foams should be defined through refining the partition function on a fixed lattice \cite{Dittrich:2012jq, Bahr:2014qza, Dittrich:2014ala}, then the melonic diagrams can be excluded by definition and one can focus on possible phase transitions which are already indicated by the distinct behaviors of the model in different ranges of $\eta$.   If one takes the point of view that the continuum limit of the model should be defined through summing over all possible diagrams corresponding to the same boundary (such as in group field theory and colored tensor models), then the melonic dominance is more worrisome. One way to resolve it is by restricting the space of diagrams which are to be summed over, and exclude the geometrically degenerate cases.  Hence future research should address a complete classification of 2-complexes which are dual to non-degenerate 4-d geometries. Another way to resolve the melonic dominance comes from the lesson in 3-d gravity. 
 In Ponzano-Regge model, one can identify a vertex translation symmetry, which is a residue of diffeomorphism symmetry resulting from discretization. Without a proper Faddeev-Popov procedure, the most divergent diagrams in the model are also melonic.  Hence we expect that in 4-d, there might be some residual gauge symmetry with non-compact gauge orbits, which could be the origin of the divergence in the 5--1 move.  Properly identifying and fixing this residual gauge symmetry might completely remove the divergence and change the behavior of melonic dominance.

\acknowledgments
I would like to thank Andrzej Banburski, Laurent Freidel, Bianca Dittrich, Aldo Riello, Jeff Hnybida and Lee Smolin for discussion and comments. I especially acknowledge Andrzej Banburski pointing me to the asymptotic formula for hypergeometrical functions, Laurent Freidel pointing torwards using the degree of graph to extract physical insights. I am very grateful to Aldo Riello and Razvan Gurau for helping me to understand the results in the colored tensor models. Research at Perimeter Institute is supported by the Government of Canada through Industry Canada and by the Province of Ontario through the Ministry of Research and Innovation. This research was also partly supported by grants from NSERC and the John Templeton
Foundation.
\appendix

\section{The loop identity}
\label{The Loop Identity}

\be
\begin{split}
&\raisebox{-12mm}{\includegraphics[keepaspectratio = true, scale = 0.45] {loopidentity.pdf}}\ \ \ \  = \ \ \ \  \sum^\infty_{A,B,J,J'=0}\!\!\!\!   \frac{N\left(A,B, J, J',\rho\right) }{A! B! J! J'!} \times \\
& \times \underbrace{ \left(\sum_{i=1}^3 [\tilde{z}^1_i|\tilde{w}^1_i\ket \right)^A }_{GF 1}  \ \ \underbrace{\left(\sum_{i=1}^3 [\tilde{z}^2_i|\tilde{w}^2_i\ket \right)^B}_{GF2}  \ \ \underbrace{\left(\sum_{i=1}^3 [z_i|w_i\ket \right)^J} _\text{Trival projection} \ \ \underbrace{\left(  \sum_{i<j<4}[z_i|z_j\ket [w_i|w_j\ket \right)^{J'}\! \!}_\text{Mixing terms},
\end{split}
\label{fullloop}
\ee
with the coefficient $N\left(A,B, J, J',\rho\right)$ given by
\be
\begin{split}
N\left(A,B, J, J',\rho\right) \equiv & \sum_{K=0}^{J'}   \frac{ J'!(J\!+K)!(J\!+\!2K\!+\!1)^\eta  }{K! (J'\!-\!K)! (J\!+\!J'\!+\!K\!+\!1)! } \frac{(-1)^{K}}{(1+\rho^2)^{(A+B +12K+7J+2J')} }  \times \\
&\times F_\rho^2\left(J+J'+K\right)  F_\rho^2\left( (A+J)/2+K\right) F_\rho^2\left( (B+J)/2+K\right),\nonumber  
\end{split}
\ee
where we have defined  $ F_\rho(J)  \equiv  {}_2F_1(-J-1,-J;2;\rho^4)$. The variables $|\tilde{z}^1_i\ket, |\tilde{w}^1_i\ket$ appear in the strands attached to the first gauge fixing term,  similarly $|\tilde{z}^2_i\ket, |\tilde{w}^2_i\ket$ appear in the second gauge fixing, while $|z_i\ket, |w_i\ket$ are are labelled for the strands we haven't gauge fixed.

\section{The asymptotics of certain types of hypergeometrical functions and modified Bessel functions}
\label{appenasy}

The hypergeometric function has asymptotic formula \cite{Handbook of Mathematical Functions} :
\begin{flalign*}
{}_2F_1(a+\lambda, \ b-\lambda; \  c ; &\  \frac{1}{2}- \frac{1}{2} u) = \\
 2^{(a+b-1)/2} &\sqrt{\zeta \sinh \zeta} \  (\lambda  + \frac{a}{2} -\frac{b}{2})^{1-c}\ \frac{ (u+1)^{(c-a-b-1)/2} } {(u-1)^{c/2} } \times\\
\times [ I_{c-1} &\left(\zeta (\lambda + \frac{a}{2}-\frac{b}{2})\right)  (1+\mathcal{O} (\lambda^{-2})) +\\
&+\frac{I_{c-2}\left(\zeta (\lambda  +a/2-b/2)\right)}{2\lambda +a-b}\  ( (c-\frac{1}{2})\ (c-\frac{3}{2}) (\frac{1}{\zeta}-\coth\zeta)+ \\  
&\ \ \ \ \ \ \ +\frac{1}{2} (2c-a-b-1)(a+b-1) \tanh\frac{\zeta}{2}  +\mathcal{O} (\lambda^{-2}) ) ],\\
&\ \ \ \ \ \ \ \ \ \ \ \ \ \ \text{for}  \  |\text{arg}\ (u-1)| <\pi ,\ \lambda \rightarrow \infty \numberthis
\end{flalign*}
where $\zeta \equiv \cosh^{-1} u$.\\  \\
Together with Pfaff transformation:
\be
 {}_2F_1(a, b; c ; u)=(1-u)^{-b}\cdot  {}_2F_1(b, c-a; c ; \frac{u}{u-1})
\ee
we can get that for $0< \rho<1$,
\begin{flalign*}
 {}_2F_1(-J -1, & -J; 2 ; \rho^4) = \\
&\frac{(1-\rho^4)^{J+3/2} \ }{4\sqrt2\  \rho^5\ \zeta^{1/2}\  (3+2J)^2 }\  [  I_1 \left((\frac{3}{2}+J)\zeta \right) \cdot(8 \zeta \ \rho^2 \ (3 + 2 J) +\mathcal{O} (J^{-2}) ) - \\
&\ \ \ \ \ \ \ \ \ -3  I_0 \left((\frac{3}{2}+J)\zeta\right) \cdot (\zeta\  (1+\rho^4)  -2 \rho^2   +\mathcal{O} (J^{-2})  ) ], \ \ \ \ \ \ \  J\rightarrow \infty \numberthis
\label{asyshyper}
\end{flalign*}
where
\be
\zeta \equiv \cosh^{-1} \left(\frac{1 + \rho^4}{1 - \rho^4}\right)
\ee
The asymptotic expansion of hypergeometric function envolves the modified Bessel function of the first kind\cite{Handbook of Mathematical Functions} :
\begin{equation}
 I_\nu (\omega) = \sum^{\infty}_{k=0}  \frac{(\frac{1}{2} \omega)^{2k+\nu}}{k! \ \Gamma(\nu + k + 1)}, \ \ \omega \in \C
\end{equation}
which is one of the two linearly independent solutions of the modified Bessel's Equation.  $I_\nu (\omega)$ enjoys the asymptotic expansion:
\begin{equation}
I_\nu (\omega) =\frac{e^\omega}{\sqrt {2 \pi \omega}} \left(  1-\frac{4 \nu^2 -1}{8 \omega}  +\mathcal{O}  (\omega^{-2} ) \right), \text{for}  \  |\text{arg}\ \omega| <\frac{\pi}{2} ,\ |\omega| \rightarrow \infty
\label{asysbessel}
\end{equation}
Thus Eq.(\ref{asyshyper}) together with Eq.(\ref{asysbessel}), we get 
\begin{flalign*}
 {}_2F_1(-J &-1, -J; 2 ; \rho^4) =\\
&\frac{e^{\left(\frac{3}{2}+J\right) \zeta } \  (1-\rho ^4)^{\frac{3}{2}+J} }{16 \sqrt{2 \pi } (3+2 J)^{7/2} \zeta ^2 \rho ^5 } \ \cdot  [6 \rho ^2-3 \zeta  \left(1+\rho ^4\right)-\\
&\ \ \ \ \ \ \ \ \ \ \ \ \ \ \ \ \ \ \ \ \ \ \ \ \ -4 \zeta ^2  (3+2 J) \left(3+3 \rho ^4-24 \rho ^2-16\  J \rho ^2\right)+\mathcal{O} (J^{-2}) ].
\end{flalign*}
After removing some constant terms due to $J$ being large, and simplifying the expression, we finally arrive at 
\be
 {}_2F_1(-J -1, -J; 2 ; \rho^4) \sim \frac{e^{\left(\frac{3}{2}+J\right) \zeta }\cdot \left(1-\rho ^4\right)^{\frac{3}{2}+J}}{2\sqrt{\pi }\  \rho ^3 \cdot J^{3/2} }, \ \ \text{for}\ \  0< \rho<1 \ \text{as}\ \  J \rightarrow \infty ,
\ee
where $\zeta \equiv \cosh^{-1}\left[ (1 + \rho^4)/(1 - \rho^4) \right]$.

For $\rho =1$, the asymptotics are much simpler, due to the fact that ${}_2F_1(-J -1, -J; 2 ; 1)$ has very simple factorial expression:
\be
{}_2F_1(-J -1, -J; 2 ; 1)= \frac{(2J+2)!}{(J+2)! \ (J+1)!}.
\ee
With Stirling's approximation of factorials 
\be
n! \sim \sqrt{2 \pi n } \left(\frac{n}{e} \right)^n  \text{as}\ \  n \rightarrow \infty
\ee
we have 
\be
{}_2F_1(-J -1, -J; 2 ; 1) \sim \frac{4^{J+1} e }{\sqrt{\pi } \cdot J^{3/2}},  \ \   \text{as}\ \ J \rightarrow \infty.
\ee

\section{The 5--1 move as an example of truncated bulk amplitude with different gauge fixing tree}
\label{51move}
In this appendix, we will use 5--1 Pachner move as an example to illustrate that using different choice of gauge fixing trees to compute truncated bulk amplitude gives rise to different result. Both of the results have the same structure as in Eq.(\ref{tbageneral}), the differences lie in the sets\footnote{Formally, these have to be multisets because a single element can appear multiple times.} $\Phi$, $\Theta$ and the $N_l$ which essentially keep track of the number of elements in the sets.

A class of optimal spanning trees in the 5-1 cable diagram is taking any of the five vertices as root, and  the four edges connected to this vertex as branches. For example, in Fig.\ref{51}, the spanning tree is $AE \cup BE \cup CE \cup DE$.  There are $|C_T |=|E|-|V| +1 = 6$ fundamental cycles.  All of them are single-strand loops and contain one original propagator per loop. Applying six corresponding truncated loop identities, we graphically arrive at the simplified diagram. Thus we can read out the truncated bulk amplitude

\begin{figure} [h]
\centering
\begin{minipage}{7.2 cm}
\includegraphics[width=1 \textwidth]{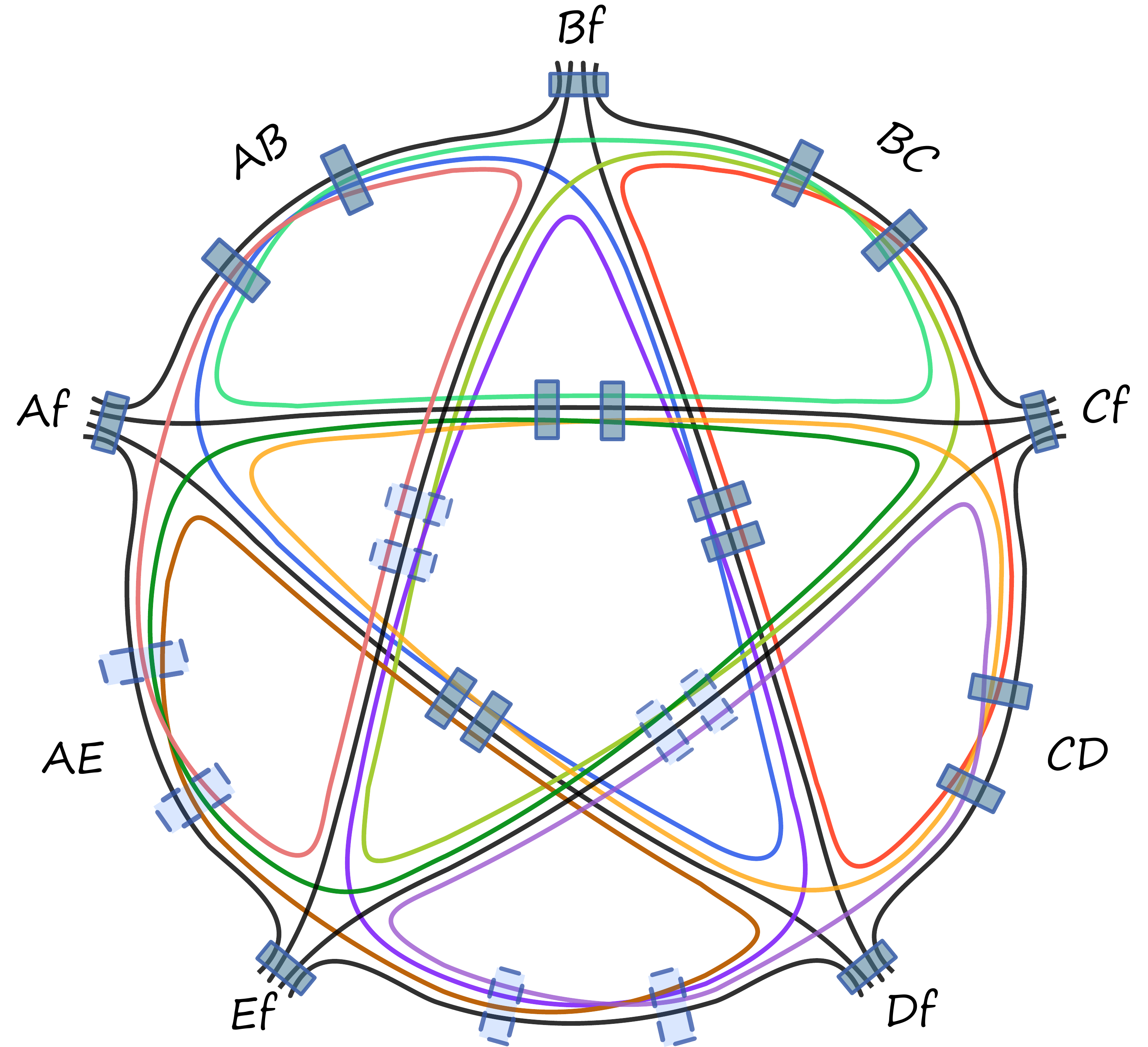}
\end{minipage}
\ \ \ \ \ \ \ \ \ \ 
\begin{minipage}{7.2 cm}
\includegraphics[width=1 \textwidth]{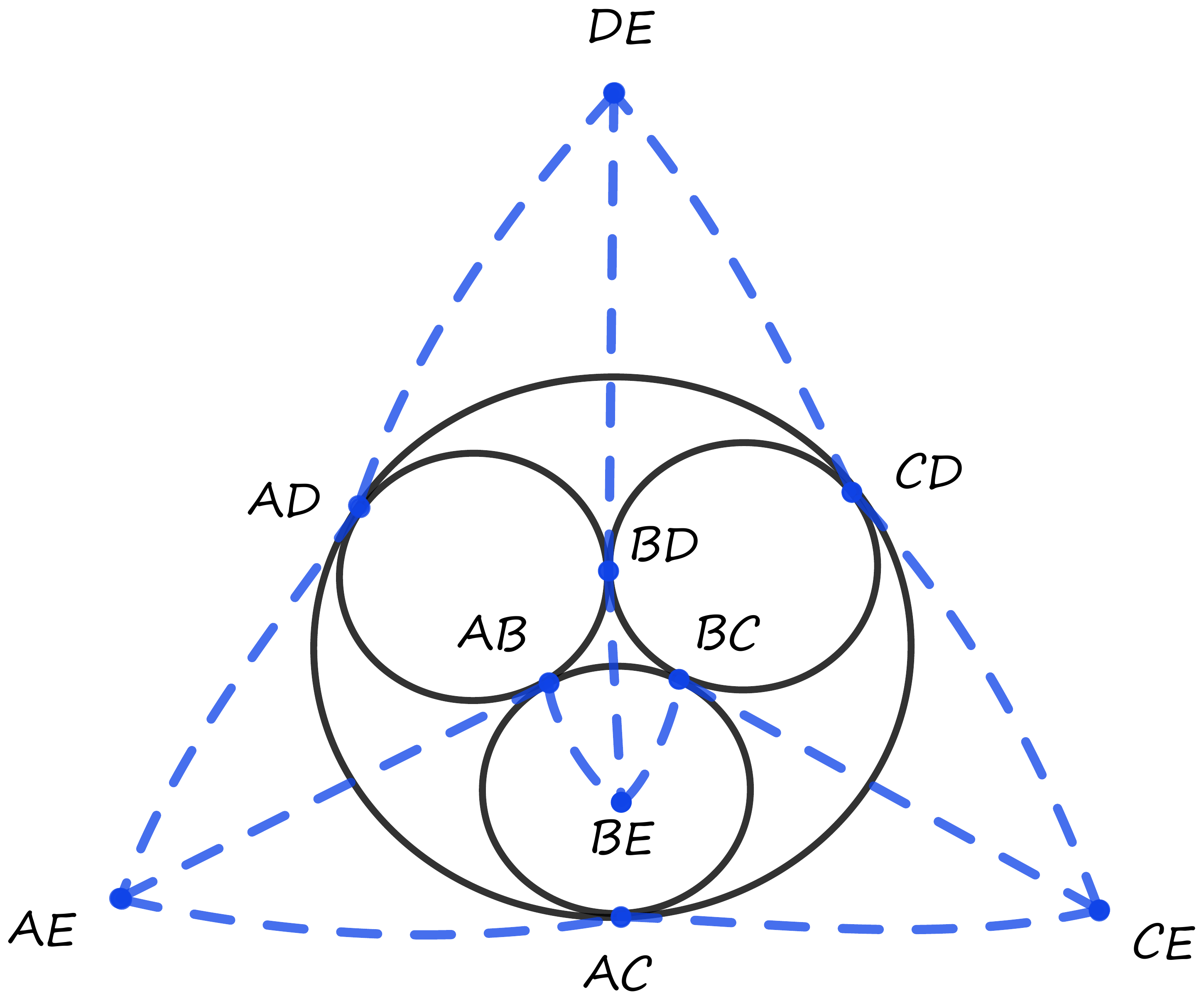}
\end{minipage}
\caption{ The cable diagram of 5--1 Pachner move and simplified diagram. The optimal tree contains four branches and there are $|C_T|=6$ fundamental cycles. The four gauge-fixed propagators are represented as 4 blue dots in the simplified diagram. After applying the cooresponding 6 truncated loop identities there are $|F|-|C_T|=4$ residual loops. Each loop idetity creates 2 non-local connections (the dashed lines  in the simplified diagram) with the 6 shared points, which correspond to 6 original propagators.  }
\label{51}
\end{figure}

\be
\!\!\!\!\!\! A_{bulk} = \sum_{\{j_l \in Z\!\!\!\!Z /2 \}} \prod_{l} \frac{(2j_l +1)^{\eta+1}}{(1+\rho^2)^{2j_l N_l}}\cdot \prod_{\alpha=1}^{|C_T|}  \left[F_{\rho}^2\ ( \sum_{ l\in \Phi_\alpha} 2j_l )\cdot (\sum_{l\in \Phi_\alpha } 2 j_l +1 )^{\eta -1} \right] \cdot \prod_{k=1}^{|V|-1} \left[ F_{\rho}^2\ ( \sum_{l\in\Theta_k } 2 j_l) \right]
\label{51}
\ee
where $N_l \equiv 4 n_f + 2 \sum_{\alpha=1} ^{n_l} \tilde{n}_\alpha =4\times 3+ 2 \times 6 =24$
\be
\begin{split}
&\Phi_{AB} = \{ABD, ABC\}, \Phi_{AD} = \{ABD, ACD\}, \Phi_{AC} = \{ ABC, ACD\}, \\
&\Phi_{BD} = \{ ABD, ABC\}, \Phi_{BC} = \{ ABC, BCD\},  \Phi_{CD} = \{ BCD, ACD\}
\end{split}
\ee
\be
\begin{split}
&\Theta_{AE} = \{ ABD, ABC, ACD \},\ \ \  \Theta_{BE} = \{ ABD, ABC, BCD \}  \\
&\Theta_{CE} = \{  ABC, BCD, ACD \},\ \ \  \Theta_{DE} = \{ ABD, BCD, ACD \}
\end{split}
\ee

Now let us redo the caculation with another spanning tree: $AB \cup AE\cup DE \cup CD$.  This is not an optimal tree, i.e. some fundamental cycles are not single loops formed by strands. For example: adding the branch $BC$ into the spanning tree creates a cycle $ABCDE \in C_T$, however, $ABCDE \notin F$. It is the same for the edges $AC, BD$: both of their cycles $ABDE, ACDE \notin F$. 

This type of gauge fixing structure leads to nesting of loop identities. For example, we can apply truncated loop identity to loop $ADE$.  Without any mixing of strands due to the truncation, we can apply truncated loop identity again to loop $ACD$. The truncation within truncation leads to a worse approximation compared with the optimal tree above. The final result of truncated bulk amplitude has the same structure as Eq.(\ref{51}), but the summation of spins in the hypergeometrical functions are different.  

\be
\begin{split}
& \Phi_{AC} = \{ ABC, AEC\},  \ \  \Phi_{AD} = \{ ABC, AEC, ABD\},  \\
&\Phi_{BC} = \{ BCD, ABC\}, \ \ \Phi_{BD} = \{ ABD, BCD\}, \\
&\Phi_{BE} = \{BCD, BCD, ABC, ABD\}, \ \ \Phi_{CE} = \{BCD, ABC, AEC\}
\end{split}
\ee

\be
\begin{split}
&\Theta_{AB} = \{BCD, ABC, ABD \},\ \ \  \Theta_{AE} = \{ BCD, ABC, AEC, ABD \}  \\
&\Theta_{CD} = \{ BCD, ABC, AEC,\},\ \ \  \Theta_{DE} = \{ BCD, ABC, AEC, ABD\}
\end{split}
\ee
We can count the occurance of each residual loop in the above sets.  We find that $N_{BCD}=N_{ABC}=36, \ N_{AEC}=N_{ABD}=24 $. One can numerically check that the value of truncted amplitude with a choice of arbitrary spanning tree is smaller than the value corresponding  to the optimal tree. This is due to the fact that the nesting of loop identities makes the truncation a worse approximation. As we have shown in Section \ref{degree}, when the amplitude is divergent this difference disappears. The degree of divergence is a tree-independent quantity.



\begin{thebibliography}{99}

\bibitem{Rovelli:2011eq} 
  C.~Rovelli,
  ``Zakopane lectures on loop gravity,''
  PoS QGQGS {\bf 2011}, 003 (2011)
  [arXiv:1102.3660 [gr-qc]].

\bibitem{Perez:2012wv} 
  A.~Perez,
  ``The Spin Foam Approach to Quantum Gravity,''
  Living Rev.\ Rel.\  {\bf 16}, 3 (2013)
  [arXiv:1205.2019 [gr-qc]].

\bibitem{Rovelli:2014ssa} 
  C.~Rovelli and F.~Vidotto,
  ``Covariant Loop Quantum Gravity : An Elementary Introduction to Quantum Gravity and Spinfoam Theory,''





\bibitem{Perini:2008pd} 
  C.~Perini, C.~Rovelli and S.~Speziale,
  ``Self-energy and vertex radiative corrections in LQG,''
  Phys.\ Lett.\ B {\bf 682}, 78 (2009)
  [arXiv:0810.1714 [gr-qc]].

\bibitem{Krajewski:2010yq} 
  T.~Krajewski, J.~Magnen, V.~Rivasseau, A.~Tanasa and P.~Vitale,
  ``Quantum Corrections in the Group Field Theory Formulation of the EPRL/FK Models,''
  Phys.\ Rev.\ D {\bf 82}, 124069 (2010)
  doi:10.1103/PhysRevD.82.124069
  [arXiv:1007.3150 [gr-qc]].




\bibitem{Riello:2013bzw} 
  A.~Riello,
  ``Self-energy of the Lorentzian Engle-Pereira-Rovelli-Livine and Freidel-Krasnov model of quantum gravity,''
  Phys.\ Rev.\ D {\bf 88}, no. 2, 024011 (2013)
  [arXiv:1302.1781 [gr-qc]].






\bibitem{Matteo1} 
  V.~Bonzom and M.~Smerlak,
  ``Bubble divergences from cellular cohomology,''
  Lett.\ Math.\ Phys.\  {\bf 93}, 295 (2010)
  [arXiv:1004.5196 [gr-qc]].
  
  \bibitem{Matteo2} 
  V.~Bonzom and M.~Smerlak,
  ``Bubble divergences: sorting out topology from cell structure,''
  Annales Henri Poincare {\bf 13}, 185 (2012)
  [arXiv:1103.3961 [gr-qc]].







\bibitem{Freidel:2009hd} 
  L.~Freidel, R.~Gurau and D.~Oriti,
  ``Group field theory renormalization - the 3d case: Power counting of divergences,''
  Phys.\ Rev.\ D {\bf 80}, 044007 (2009)
  [arXiv:0905.3772 [hep-th]].


\bibitem{Carrozza:2011jn} 
  S.~Carrozza and D.~Oriti,
  ``Bounding bubbles: the vertex representation of 3d Group Field Theory and the suppression of pseudo-manifolds,''
  Phys.\ Rev.\ D {\bf 85}, 044004 (2012)
  doi:10.1103/PhysRevD.85.044004
  [arXiv:1104.5158 [hep-th]].

\bibitem{Baratin:2014yra} 
  A.~Baratin, L.~Freidel and R.~Gurau,
  ``Weighting bubbles in group field theory,''
  Phys.\ Rev.\ D {\bf 90}, no. 2, 024069 (2014)
  doi:10.1103/PhysRevD.90.024069
  [arXiv:1405.2808 [hep-th]].

\bibitem{Carrozza:2012kt} 
  S.~Carrozza and D.~Oriti,
  ``Bubbles and jackets: new scaling bounds in topological group field theories,''
  JHEP {\bf 1206}, 092 (2012)
  doi:10.1007/JHEP06(2012)092
  [arXiv:1203.5082 [hep-th]].



\bibitem{Baratin:2013rja} 
  A.~Baratin, S.~Carrozza, D.~Oriti, J.~Ryan and M.~Smerlak,
  ``Melonic phase transition in group field theory,''
  Lett.\ Math.\ Phys.\  {\bf 104}, 1003 (2014)
  doi:10.1007/s11005-014-0699-9
  [arXiv:1307.5026 [hep-th]].



\bibitem{Freidel:2002dw} 
  L.~Freidel and D.~Louapre,
  ``Diffeomorphisms and spin foam models,''
  Nucl.\ Phys.\ B {\bf 662}, 279 (2003)
  [gr-qc/0212001].



\bibitem{Baratin:2006gy} A.~Baratin and L.~Freidel, 
``Hidden Quantum Gravity in 4-D Feynman diagrams: Emergence of spin foams,''
  Class.\ Quant.\ Grav.\ {\bf 24}, 2027 (2007)
  doi:10.1088/0264-9381/24/8/007
  [hep-th/0611042].



\bibitem{Bonzom:2013ofa} 
  V.~Bonzom and B.~Dittrich,
  ``Bubble divergences and gauge symmetries in spin foams,''
  Phys.\ Rev.\ D {\bf 88}, 124021 (2013)
  doi:10.1103/PhysRevD.88.124021
  [arXiv:1304.6632 [gr-qc]].



\bibitem{Dittrich:2011vz} 
  B.~Dittrich and S.~Steinhaus,
  ``Path integral measure and triangulation independence in discrete gravity,''
  Phys.\ Rev.\ D {\bf 85}, 044032 (2012)
  [arXiv:1110.6866 [gr-qc]].


\bibitem{Dittrich:2014rha} 
  B.~Dittrich, W.~Kaminski and S.~Steinhaus,
  ``Discretization independence implies non-locality in 4D discrete quantum gravity,''
  arXiv:1404.5288 [gr-qc].




\bibitem{Dittrich:2008pw} 
  B.~Dittrich,
  ``Diffeomorphism symmetry in quantum gravity models,''
  Adv.\ Sci.\ Lett.\  {\bf 2}, 151
  doi:10.1166/asl.2009.1022
  [arXiv:0810.3594 [gr-qc]].



\bibitem{Bahr:2009ku} 
  B.~Bahr and B.~Dittrich,
  ``(Broken) Gauge Symmetries and Constraints in Regge Calculus,''
  Class.\ Quant.\ Grav.\  {\bf 26}, 225011 (2009)
  doi:10.1088/0264-9381/26/22/225011
  [arXiv:0905.1670 [gr-qc]].




%





\bibitem{Oeckl:2004yf} 
  R.~Oeckl,
  ``Renormalization for spin foam models of quantum gravity,''
  In *Rio de Janeiro 2003, Recent developments in theoretical and experimental general relativity, gravitation, and relativistic field theories, pt. C* 2296-2300
  [gr-qc/0401087].

\bibitem{Oeckl:2002ia} 
  R.~Oeckl,
  ``Renormalization of discrete models without background,''
  Nucl.\ Phys.\ B {\bf 657}, 107 (2003)
  [gr-qc/0212047].


\bibitem{Bahr:2011uj} 
  B.~Bahr, B.~Dittrich and S.~Steinhaus,
  ``Perfect discretization of reparametrization invariant path integrals,''
  Phys.\ Rev.\ D {\bf 83}, 105026 (2011)
  doi:10.1103/PhysRevD.83.105026
  [arXiv:1101.4775 [gr-qc]].


\bibitem{Dittrich:2012jq} 
  B.~Dittrich,
  ``From the discrete to the continuous: Towards a cylindrically consistent dynamics,''
  New J.\ Phys.\  {\bf 14}, 123004 (2012)
  [arXiv:1205.6127 [gr-qc]].

\bibitem{Bahr:2014qza} 
  B.~Bahr,
  ``On background-independent renormalization of spin foam models,''
  arXiv:1407.7746 [gr-qc].

\bibitem{Dittrich:2014ala} 
  B.~Dittrich,
  ``The continuum limit of loop quantum gravity - a framework for solving the theory,''
  arXiv:1409.1450 [gr-qc].








\bibitem{Gurau:2010ba} 
  R.~Gurau,
  ``The 1/N expansion of colored tensor models,''
  Annales Henri Poincare {\bf 12}, 829 (2011)
  doi:10.1007/s00023-011-0101-8
  [arXiv:1011.2726 [gr-qc]].

\bibitem{Gurau:2011xq} 
  R.~Gurau,
  ``The complete 1/N expansion of colored tensor models in arbitrary dimension,''
  Annales Henri Poincare {\bf 13}, 399 (2012)
  doi:10.1007/s00023-011-0118-z
  [arXiv:1102.5759 [gr-qc]].


\bibitem{Gurau:2011aq} 
  R.~Gurau and V.~Rivasseau,
  ``The 1/N expansion of colored tensor models in arbitrary dimension,''
  Europhys.\ Lett.\  {\bf 95}, 50004 (2011)
  doi:10.1209/0295-5075/95/50004
  [arXiv:1101.4182 [gr-qc]].


\bibitem{Bonzom:2011zz} 
  V.~Bonzom, R.~Gurau, A.~Riello and V.~Rivasseau,
  ``Critical behavior of colored tensor models in the large N limit,''
  Nucl.\ Phys.\ B {\bf 853}, 174 (2011)
  doi:10.1016/j.nuclphysb.2011.07.022
  [arXiv:1105.3122 [hep-th]].

\bibitem{Gurau:2011xp} 
  R.~Gurau and J.~P.~Ryan,
  ``Colored Tensor Models - a review,''
  SIGMA {\bf 8}, 020 (2012)
  doi:10.3842/SIGMA.2012.020
  [arXiv:1109.4812 [hep-th]].





\bibitem{Gurau:2013cbh} 
  R.~Gurau and J.~P.~Ryan,
  ``Melons are branched polymers,''
  Annales Henri Poincare {\bf 15}, no. 11, 2085 (2014)
  doi:10.1007/s00023-013-0291-3
  [arXiv:1302.4386 [math-ph]].




\bibitem{Gabrielli:1997zy} 
  D.~Gabrielli,
  ``Polymeric phase of simplicial quantum gravity,''
  Phys.\ Lett.\ B {\bf 421}, 79 (1998)
  doi:10.1016/S0370-2693(98)00022-7
  [hep-lat/9710055].









%
%
%
%
%




%
%
%
%

%
%
%
%
%






\bibitem{Banburski:2014cwa} 
  A.~Banburski, L.~Q.~Chen, L.~Freidel and J.~Hnybida,
  ``Pachner moves in a 4d Riemannian holomorphic Spin Foam model,''
  Phys.\ Rev.\ D {\bf 92}, no. 12, 124014 (2015)
  doi:10.1103/PhysRevD.92.124014
  [arXiv:1412.8247 [gr-qc]].


\bibitem{Banburski:2015kmc} 
  A.~Banburski and L.~Q.~Chen,
  ``A simpler way of imposing simplicity constraints,''
  arXiv:1512.05331 [gr-qc].



\bibitem{coh1}
E.R. Livine and S. Speziale,
``A new spinfoam vertex for quantum gravity",
Phys.Rev.D76 (2007) 084028 [arXiv:0705.0674]


\bibitem{Livine:2007ya} 
  E.~R.~Livine and S.~Speziale,
  ``Consistently Solving the Simplicity Constraints for Spinfoam Quantum Gravity,''
  Europhys.\ Lett.\  {\bf 81}, 50004 (2008)
  [arXiv:0708.1915 [gr-qc]].



\bibitem{Freidel:2009nu} 
  L.~Freidel, K.~Krasnov and E.~R.~Livine,
  ``Holomorphic Factorization for a Quantum Tetrahedron,''
  Commun.\ Math.\ Phys.\  {\bf 297}, 45 (2010)
  [arXiv:0905.3627 [hep-th]].



\bibitem{Freidel:2009ck} 
  L.~Freidel and E.~R.~Livine,
  ``The Fine Structure of SU(2) Intertwiners from U(N) Representations,''
  J.\ Math.\ Phys.\  {\bf 51}, 082502 (2010)
  [arXiv:0911.3553 [gr-qc]].

\bibitem{Freidel:2010tt} 
  L.~Freidel and E.~R.~Livine,
  ``U(N) Coherent States for Loop Quantum Gravity,''
  J.\ Math.\ Phys.\  {\bf 52}, 052502 (2011)
  [arXiv:1005.2090 [gr-qc]].


\bibitem{Dupuis:2010iq} 
  M.~Dupuis and E.~R.~Livine,
  ``Revisiting the Simplicity Constraints and Coherent Intertwiners,''
  Class.\ Quant.\ Grav.\  {\bf 28}, 085001 (2011)
  [arXiv:1006.5666 [gr-qc]].

\bibitem{Borja:2010rc} 
  E.~F.~Borja, L.~Freidel, I.~Garay and E.~R.~Livine,
  ``U(N) tools for Loop Quantum Gravity: The Return of the Spinor,''
  Class.\ Quant.\ Grav.\  {\bf 28}, 055005 (2011)
  [arXiv:1010.5451 [gr-qc]].



\bibitem{Dupuis:2011fz} 
  M.~Dupuis and E.~R.~Livine,
  ``Holomorphic Simplicity Constraints for 4d Spinfoam Models,''
  Class.\ Quant.\ Grav.\  {\bf 28}, 215022 (2011)
  [arXiv:1104.3683 [gr-qc]].



\bibitem{Livine:2011gp} 
  E.~R.~Livine and J.~Tambornino,
  ``Spinor Representation for Loop Quantum Gravity,''
  J.\ Math.\ Phys.\  {\bf 53}, 012503 (2012)
  [arXiv:1105.3385 [gr-qc]].


\bibitem{Dupuis:2011wy} 
  M.~Dupuis, L.~Freidel, E.~R.~Livine and S.~Speziale,
  ``Holomorphic Lorentzian Simplicity Constraints,''
  J.\ Math.\ Phys.\  {\bf 53}, 032502 (2012)
  [arXiv:1107.5274 [gr-qc]].



\bibitem{Dupuis:2012vp} 
  M.~Dupuis, S.~Speziale and J.~Tambornino,
  ``Spinors and Twistors in Loop Gravity and Spin Foams,''
  PoS QGQGS {\bf 2011}, 021 (2011)
  [arXiv:1201.2120 [gr-qc]].




\bibitem{Freidel:2012ji} 
  L.~Freidel and J.~Hnybida,
  ``On the exact evaluation of spin networks,''
  arXiv:1201.3613 [gr-qc].



\bibitem{Freidel:2013fia} 
  L.~Freidel and J.~Hnybida,
  ``A Discrete and Coherent Basis of Intertwiners,''
  arXiv:1305.3326 [math-ph].


\bibitem{Hnybida:2014mwa} 
  J.~Hnybida,
  ``Generating Functionals for Spin Foam Amplitudes,''
  arXiv:1411.2049 [math-ph].



\bibitem{Freidel:2005qe} 
  L.~Freidel,
  ``Group field theory: An Overview,''
  Int.\ J.\ Theor.\ Phys.\  {\bf 44}, 1769 (2005)
  doi:10.1007/s10773-005-8894-1
  [hep-th/0505016].


\bibitem{Oriti:2006se} 
  D.~Oriti,
  ``The Group field theory approach to quantum gravity,''
  In *Oriti, D. (ed.): Approaches to quantum gravity* 310-331
  [gr-qc/0607032].





\bibitem{Conrady:2008mk} 
  F.~Conrady and L.~Freidel,
  ``On the semiclassical limit of 4d spin foam models,''
  Phys.\ Rev.\ D {\bf 78}, 104023 (2008)
  [arXiv:0809.2280 [gr-qc]].

\bibitem{Barrett:2009gg} 
  J.~W.~Barrett, R.~J.~Dowdall, W.~J.~Fairbairn, H.~Gomes and F.~Hellmann,
  ``Asymptotic analysis of the EPRL four-simplex amplitude,''
  J.\ Math.\ Phys.\  {\bf 50}, 112504 (2009)
  [arXiv:0902.1170 [gr-qc]].

\bibitem{Han:2011rf} 
  M.~X.~Han and M.~Zhang,
  ``Asymptotics of Spinfoam Amplitude on Simplicial Manifold: Euclidean Theory,''
  Class.\ Quant.\ Grav.\  {\bf 29}, 165004 (2012)
  [arXiv:1109.0500 [gr-qc]].

\bibitem{Regge:1961px} 
  T.~Regge,
  ``General Relativity Without Coordinates,''
  Nuovo Cim.\  {\bf 19}, 558 (1961).




\bibitem{Bargmann}
  V.~Bargmann,
  ``On the Representations of the Rotation Group,''
  Rev.\ Mod.\ Phys.\  {\bf 34}, 829 (1962).
	

\bibitem{Schwinger}
  J.~Schwinger,
  ``On Angular Momentum,''
  U.S. \ Atomic \ Energy \ Commission.\ (unpublished) \ NYO-3071, (1952).	


\bibitem{Kocay}
 W.~Kocay, D.L.~Kreher, ``Graphs, Algorithms, and Optimization, Discrete Mathematics and Its Applications", CRC Press (2004).



\bibitem{Freidel:2002xb} 
  L.~Freidel and E.~R.~Livine,
  ``Spin networks for noncompact groups,''
  J.\ Math.\ Phys.\  {\bf 44}, 1322 (2003)
  [hep-th/0205268].






\bibitem{Handbook of Mathematical Functions} 
F. W. J. Olver, D. W. Lozier, R. F. Boisvert, and C. W. Clark, editors. 
NIST Handbook of Mathematical Functions. 
Cambridge University Press, New York, NY, 2010. 
Print companion to [DLMF].

\end{thebibliography}
\end{document}